\documentclass[superscriptaddress,nofootinbib
]{revtex4}
\usepackage{amssymb,amsmath}
\usepackage{parcolumns}
\usepackage{url}
\usepackage{subfigure}
\usepackage{textcomp}
\usepackage{graphicx}
\usepackage{bm}
\usepackage{dcolumn}
\usepackage[usenames]{color}
\usepackage{epsfig}
\usepackage{xcolor}
\input epsf.tex
\usepackage{amssymb}
\usepackage{amsmath}
\usepackage{bm}
\usepackage{amsthm}
\usepackage{braket}
\usepackage[title]{appendix}
\usepackage{amsopn}
\usepackage[breaklinks,colorlinks,bookmarks=false,citecolor=blue,linkcolor=blue,urlcolor=blue]{hyperref}
\def\comment#1{}

\def\ket#1{\mathinner{|{#1}\rangle}}

\def\beq{\begin{equation}}
\def\eeq{\end{equation}}
\def\bea{\begin{eqnarray}}
\def\eea{\end{eqnarray}}

\makeatletter
\usepackage{ulem}
\usepackage{cancel}
\usepackage{color}

\makeatother

\begin{document}

\title{Photon self-interaction through gravitons and axions}

\author{Ahmad Hoseinpour}
\email[]{ahmadhoseinpoor454@gmail.com}

\affiliation{Department of Physics, Isfahan University of Technology, Isfahan
84156-83111, Iran}

\author{Moslem Zarei}
\email[]{m.zarei@iut.ac.ir}

\affiliation{Department of Physics, Isfahan University of Technology, Isfahan
84156-83111, Iran}

\author{Mehdi Abdi}
\email[]{mehabdi@gmail.com}
\affiliation{Wilczek Quantum Center, School of Physics and Astronomy, Shanghai Jiao Tong University, 200240 Shanghai, China}
\affiliation{Department of Physics, Isfahan University of Technology, Isfahan 84156-83111, Iran}

\date{\today}

 \begin{abstract}
In this work, we propose to employ the concept of photon self-interaction for axion detection. In particular, we derive the interaction Hamiltonian for photons via axions in a ring cavity. We show that when the incoming photons are considered in plane-wave basis, the interaction vanishes. However, when the realistic case of photon wavepackets are assumed, a self-interaction whose strength is proportional to the size of the wavepacket and the cavity length exists. Under specific conditions, we find that the axion-mediated interaction dominates the gravitationally induced self-interaction.
We discuss the implications of this setup for axion detection, focusing on the range of axion mass, $10^{-10}~\text{eV}<m_a<10^{-4}~\text{eV}$ for which the best accuracy of coupling constants is constrained in some cases to $g_{a\gamma \gamma}> 9 \times 10^{-12}~ \text{GeV}^{-1}$. 
\end{abstract}

\maketitle


\section{Introduction}
\begin{parcolumns}{2}
In recent years, quantum technologies including advanced quantum sensing, quantum metrology, and the detectors based on them have experienced extraordinary growth \cite{Kaltenbaek:2021, Carney:2021, Gatti:2021, Stray:2022, Golwala:2022, Chou:2023, Sushkov:2023, Tobar:2024, Ye:2024, DeMille:2024}. It is anticipated that pioneering experiments employing these technologies will uncover previously unknown particles, such as axions. Moreover, additional experiments have been proposed to either validate or challenge the quantum nature of gravity~ \cite{Pikovski:2012, Hogan:2012, Marin:2013, Kumar:2018, Howl:2021, Zain:2023, Marshman:2020, Bose:2017, Marletto:2017}.
The laboratory-based axion experiments \cite{Bibber:1987, Wuensch:1989, Bhre:2013, Beck:2013}, together with recent axion test proposals \cite{Graham:2011, Sikivie:2014a, Sikivie:2014b, Graham:2018, Chang:2019, Goryachev:2019, Fedderke:2019, Zarei:2022, Yavuz:2022, Yang:2023, Hajebrahimi:2023}, as well as recently proposed experiments investigating the quantum nature of gravity, have created new opportunities to search for gravitons and axions. 
It is worth noting that in some quantum gravity (QG) proposals \cite{Bose:2017, Marletto:2017, Sugiyama:2023, Sugiyama:2024}, an entanglement witness in the resulting quantum states has been introduced as a signature of QG.
In some proposals \cite{Howl:2021, Zain:2023}, the non-Gaussianity in the matter state are investigated as a signature of QG.
 Meanwhile, in some other proposed axion tests \cite{Fedderke:2019, Zarei:2022}, the change in the polarization of the photon beam is considered as a signature of detection. Given the significant advantages of non-Gaussianity signature, this article will also explore one of its related proposals.
 The importance of axion detection arises from the fact that, after resolving the strong CP problem with the QCD axion, all types of axions \cite{Peccei:1977, Weinberg:1978, Wilczek:1978, Kim:2010}, as well as axion-like particles (ALPs) \cite{Witten:2006, Jaeckel:2010, Arias:2012}, have been proposed as potential solutions to the nature of dark matter \cite{Abbott:1983, Dine:1983, Preskill:1983}.
There are several types of axion detection experiments. Resonant cavity experiments \cite{Sikivie:1983}, such as axion haloscopes and axion helioscopes, play a crucial role in this field. Axion haloscopes convert halo axions into photons in the presence of a magnetic field within a microwave cavity. Notable examples include QUAX-LNF \cite{Rettaroli:2024} and ADMX \cite{Asztalos:2010, Rybka:2014}, which are cavity haloscopes, as well as MADMAX \cite{Caldwel:2017}, a dielectric haloscope, all operating at the microwave scale.
Axion helioscopes \cite{Zioutas:1999} search for converted photons from the axions of the Sun as a natural source of these particles. The first well-known axion helioscope is the CAST, based on the conversion of solar axions back to photons in a strong magnet of $\sim 9.5~ \text{Tesla}$ \cite{Zioutas:1999, Zioutas:2005, Andriamonje:2007}. It determines the constraint $g_{a\gamma}>6.6 \times 10^{-11} \, \text{GeV}^{-1}$ for the range of axion masses below 10 meV \cite{OHare:2020}. The IAXO \cite{Irastorza:2013, Armengaud:2014} is a next-generation axion helioscope aiming to improve the sensitivity in coupling by one order of magnitude compared to CAST, to explore ALPs and to probe QCD axion models to scales below meV. The experiment OSQAR-LSW is in collaboration with CAST, looking for ALPs by examining the optical properties of the quantum vacuum consisting of a strong magnetic field using light shining through a wall \cite{Redondo:2011, Beyer:2022}. The experiments CAST-CAPP and the Relic Axion Detector Exploratory Setup (RADES) both aim to use the CAST magnet to develop an effective high-mass axion haloscope to search for DM axions. In a different axion DM experiment, low-mass axion dark matter could be searched for using cosmic microwave background polarization \cite{Fedderke:2019}, because the polarization of light rotates when passing through axion dark matter.

Advancements in tabletop quantum experiments designed to elucidate the enigmatic nature of gravity offer a remarkable opportunity to extend toward findings that support the existence of axions. Accordingly, in this work, we study a gravity tabletop proposal from a quantum field theory (QFT) perspective and generalize it such that it becomes suitable for probing axions. To investigate the realized test, we study the case of a Gaussian photon beam in addition to the simplified plane-wave beam in order to identify any potential differences.
From this work, two significant conclusions arise. The first is that studying quantum interactions within the wavepacket basis assumption yields crucially different results compared to the plane-wave case. The second is that the considered QG test proposal could also serve as an axion detector. Therefore, such QG test proposals require careful consideration of how to distinguish and separate axionic effects from gravitational ones.

In what follows, we discuss in Section \ref{S-NG} a brief review of the desired test setup and non-Gaussianity, respectively. Readers already familiar with these topics can skip this section without losing the article's flow. In Section \ref{S-GRAVITY}, we explain in detail the gravity interaction of two laser beams in the parallel arms of a ring setup, using the QFT formalism. Then, we explain the corrections by considering a Gaussian wavepacket instead of a plane-wave beam. In Section \ref{S-AXION}, we consider an axion mediator instead of a graviton because we are interested in axion detection and its properties. In the remainder of this paper, we calculate the theoretical precision of the proposed quantum setup using the Cramér-Rao bound and finally evaluate the theoretical bounds on axion parameters for the introduced ring cavity (RC) proposal.

\end{parcolumns}

\section{quantum gravity ring cavity proposal and signature of non-Gaussianity}\label{S-NG}

Studying QG tests based on quantum information resources has recently motivated much theoretical interest \cite{Bose:2017, Marletto:2017, Howl:2021, Marshman:2020, Zain:2023}. Within these QG tests, we are interested in revising \cite{Zain:2023} and subsequently exploring their generalization for axion tests.

\begin{figure}
\centering
   \includegraphics[width=6.5 in]{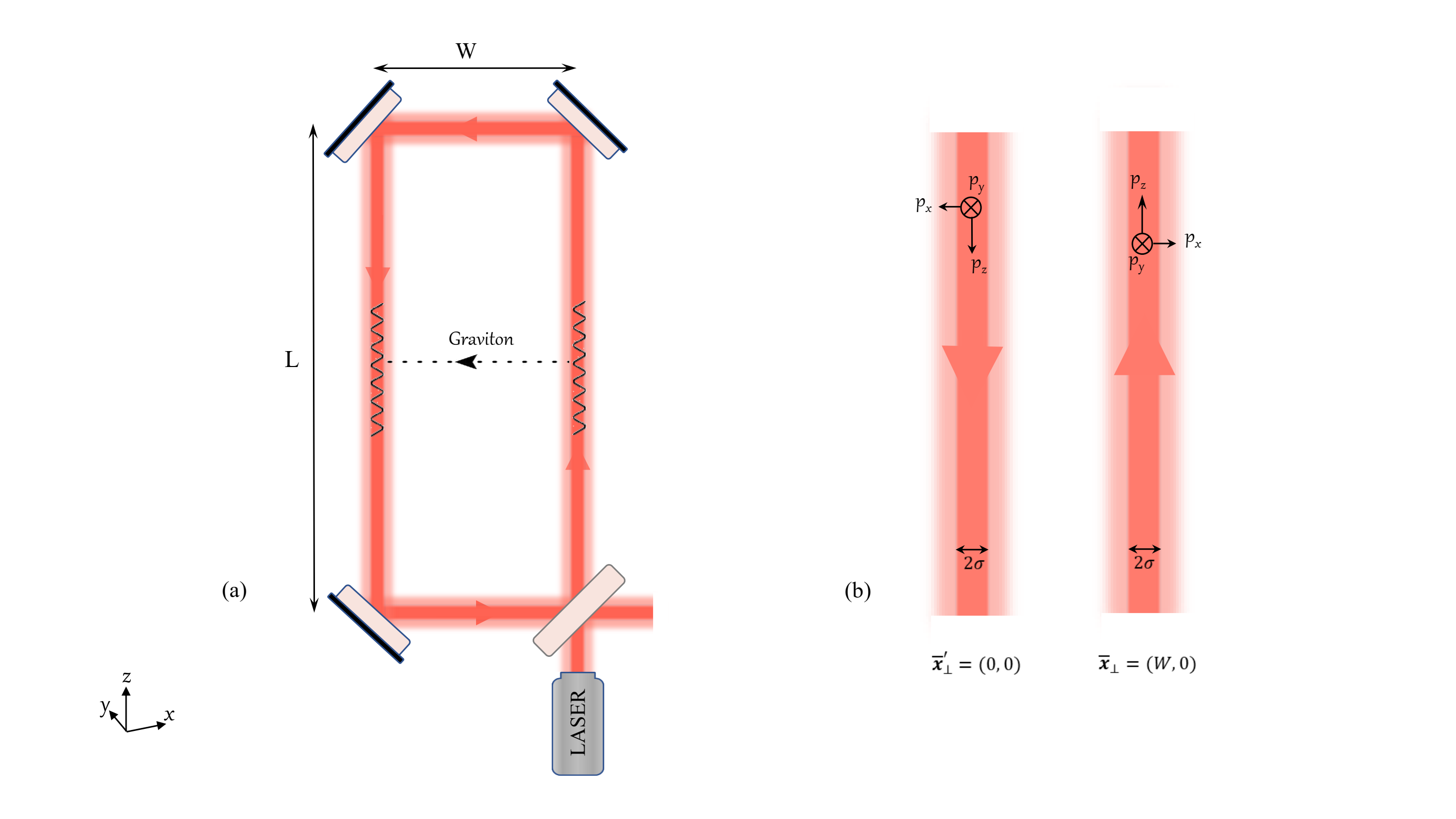}\\
  \caption{ a) A simplified optical ring cavity with the corresponding Feynman diagram for the gravitational self-interaction of photons. b) The photon momentum direction within the long arms of the cavity \cite{Zain:2023}.
 }\label{ring}
\end{figure}

In this section, we briefly review the setup proposed in \cite{Zain:2023} with a ring cavity test based on the gravitational self-interaction of photons (see Fig. \ref{ring}a). In the same framework that this paper focuses on, a laser beam propagates within a rectangular ring characterized by a length $L$ and width $W$, assuming $L \gg W$. This configuration allows us to focus primarily on the significant self-interactions of photons along the longitudinal length $L$.

In \cite{Zain:2023}, the quantum parameters are estimated using the two-parameter estimation theory. The Hamiltonian of photon self-interaction in a ring cavity (Fig. \ref{ring}a) has been obtained as \cite{Zain:2023}
\bea
\hat{H}_I=-\frac{ 16 G }{L} \left(\frac{\hbar \omega_0}{c^2}\right)^2  \ln\left( \frac{L}{W}\right)~\hat{a}^\dagger \hat{a} \hat{a}^\dagger  \hat{a}~, \label{Zain-Hamilton}
\eea
where $G $ represents the universal gravitational constant, $c$ denotes the speed of light, $\hbar$ signifies the reduced Planck constant. Additionally, $ L $ refers to the length of the cavity, $W$ indicates the width of the cavity, $\omega_0$ represents the single mode of the cavity, while $a, a^\dagger$ are the annihilation and creation operators of the photon, respectively. Subsequently, the corresponding unitary operator \( \hat{U}_{\text{Q}} \) is extracted to describe the evolution of the initial quantum state of the photons.
\bea
\ket{\psi}=\hat{U}_{\text{Q}}\ket{\psi_0}~,
\eea
where 
\bea
\hat{U}_{\text{Q}}=\text{exp}( i \chi_{\text{Q}}~ \hat{a}^\dagger \hat{a}\hat{a}^\dagger \hat{a}) ~,
\eea
and \(\ket{\psi_0}\) is the initial squeezed vacuum (SQV) state.
Next, the quantum parameter
\bea
\chi_{\text{Q}} \approx32 G\mathcal{F}\left(\frac{\hbar \omega_0^2}{c^5}\right) \ln\left(\frac{L}{W}\right)~,
\eea
which is related to the Hamiltonian multiplied by the time the photon spends in the cavity $\tau=2\mathcal{F}L/c$, has been obtained for estimation, where $\mathcal{F}$ represents the finesse of the cavity. The optimal theoretical sensitivity to this quantum parameter ($\chi_{\text{Q}}$) has been estimated using quantum Fisher information (QFI) \cite{Braunstein:1994, Braunstein:1996, Liu:2019}. Therefore, the $\chi_{\text{Q}}$ parameter estimation yields \cite{Zain:2023}
\bea
\chi_{\text{Q}} \geqslant \left[\sqrt{96M}N(N+1)\right]^{-1}~,\label{CREq}
\eea
where $N$ is the mean photon number in the cavity and $M$ represents the number of independent experiments.
Note that Eq. \eqref{CREq} is obtained using the Cramér-Rao bound \cite{Helstrom:1967, Liu:2019}, which indicates the minimum value required for $\chi_{\text{Q}}$ to detect the interaction. The cavity proposed in \cite{Zain:2023} has a length of $L = 10$ km and operates with a laser wavelength of $2 \, \mu\text{m}$. 
To detect the quantum signature of gravity over an interrogation period of one year, a minimum power of approximately 125 MW is required. This QG test necessitates advancements in experimental capabilities, including high-efficiency photon detectors, to reveal the corresponding signatures.\\
To design such QG test proposals, there are three key remarks to consider: 1) As an optical setup, it requires purely gravitational interactions between photons. Therefore, it is essential to minimize or eliminate electromagnetic interactions compared to gravitational interactions. The BMV gravity test proposal \cite{Bose:2017, Marletto:2017} achieves this by ensuring that the microparticles have no excess charge and by increasing the separation between the particles until the gravitational signal remains significant. Meanwhile, the Bose–Einstein condensate (BEC) quantum gases test \cite{Howl:2021} suppresses the electromagnetic interactions noise using an external magnetic or optical field without decreasing the strength of the gravitational interactions. For this purpose, Ref. \cite{Zain:2023} suggests employing a traveling wave in a ring cavity to render the QED interaction negligible. 2) The existence of quantum systems is crucial for ensuring purely quantum interactions of gravity. In the BMV proposals, this is achieved with spherical nanoparticles that are sufficiently small to allow for the superposition of their locations as a quantum state. In \cite{Zain:2023}, photons are employed as quantum matter to investigate their interaction with QG. Although \cite{Zain:2023} emphasizes that the benefit of the proposed test, compared to previous non-relativistic experiments, lies in offering a more detailed understanding of the true nature of the gravitational field—particularly the relativistic aspects of QG—we argue that certain neglected entities, such as axions (if they exist), could potentially produce results akin to those of gravity. Consequently, to conduct a robust test of gravity, more precise measurements and detailed filtering techniques are necessary. 3) Another consideration is selecting the most suitable QG signature for measurement, such as non-Gaussianity.

Recent works \cite{Howl:2021, Zain:2023} discuss that creating non-Gaussianity in the matter state through pure gravitational interactions would not be feasible with a classical theory of gravity, whereas a quantum theory of gravity would permit it. Therefore, non-Gaussianity, as opposed to other quantum information witnesses like quantum correlations and entanglement \cite{Bose:2017, Marletto:2017}, serves as a strong signature of QG. An additional advantage of non-Gaussianity as a witness for QG is that it requires only a single tabletop quantum system, unlike previous tests that necessitated multipartite quantum systems.\\
For instance, \cite{Howl:2021} has proposed a tabletop test of QG based on a single quantum system that utilizes a non-Gaussianity signature from a Bose-Einstein condensate (BEC) in a single well. Ref. \cite{Zain:2023} suggests probing the non-Gaussianity in the emitted photons from a single cavity. Additionally, \cite{Howl:2021} recommends using a squeezed state instead of the BMV proposal, which effectively employs a NOON state that is challenging to implement in practice.

\section {Gravitational self-interaction of photons in the ring cavity}\label{S-GRAVITY}

In this section, we rederive Hamiltonian \eqref{Zain-Hamilton} using the QFT calculations based on the setup introduced in \cite{Zain:2023}, where a rectangular ring is proposed to consider the self-interaction of photons in the cavity. We begin with the photon-graviton vertex interaction in natural units
\bea
\mathcal{L}_{g\gamma\gamma}=\frac{-\kappa}{16 \pi} \left(\frac{1}{2}h_\alpha^\alpha~F_{\mu \nu}F^{\mu \nu}-2h^{\mu \nu}~F_\mu ^\alpha F_{\alpha \nu}\right)~,
\eea
where $\kappa = \sqrt{16 \pi G}$, $ F_{\mu \nu}(x) = \partial_\mu A_\nu(x) - \partial_\nu A_\mu(x)$ is the electromagnetic field strength tensor and $h^{\mu \nu}$ is the graviton field. The Hamiltonian of the photon-graviton vertex becomes
\bea
\mathcal{H}_{g\gamma\gamma}=\frac{\kappa}{16 \pi} \left(\frac{1}{2}h_\alpha^\alpha~F_{\mu \nu}F^{\mu \nu}-2h^{\mu \nu}~F_\mu ^\alpha F_{\alpha \nu}\right)~.
\eea
Using the relation of the second-order S-matrix \cite{Kosowsky:1996}
\bea
\mathcal{S}^{(2)}&=&-\frac{1}{2}\int_{-\infty}^{\infty}dt \int_{-\infty}^{ \infty}dt' ~T\{H_{g\gamma \gamma}(t) H_{g \gamma \gamma} (t')\} = -i \int_{-\infty}^{ \infty}dt H^{(2)}(t)~,\label{s-matrix}
\eea
one can obtain the corresponding Hamiltonian, $H_I$, related to the tree-level Feynman diagram in Fig. \ref{ring}a. 
Here, the second-order S-matrix refers to interactions with two Feynman vertices, and $T$ denotes the time-ordered operator. Thus, equation \eqref{s-matrix} yields
\bea
H_I(t)&=&4\left(\frac{\kappa}{16\pi}\right)^2 \int d^4 x' \int d^3\bold{x}~  F_\mu ^\rho (x) F_{\rho \nu}(x) D_{\mu \nu \alpha \beta}(x-x')  F_\alpha ^\zeta (x') F_{\zeta \beta}(x')~, 
\eea
where
\bea
 D_{\mu \nu \alpha \beta}(x-x')=\int \frac{d^4k}{(2\pi)^4}~\frac{i e^{-ik.(x-x')}}{k^2}\left(\eta_{\mu \alpha} \eta_{\nu \beta}+\eta_{\mu \beta} \eta_{\nu \alpha}-\eta_{\mu \nu} \eta_{\alpha \beta}\right)~,\nonumber
\eea
is the graviton propagator. 
To realize the proposed setup, we reinterpret the QFT calculations by employing a wavepacket basis for the electromagnetic field, in contrast to the plane-wave expansion. Our motivation arises from the hypothesis that a theoretical consideration related to a real experiment conducted at macroscopic scales will yield accurate results when wavepacket considerations are incorporated. To this end, we utilize the photon fields in the basis of Gaussian wavepackets in our QFT calculations
\bea
A_\mu (\bold{x}) \propto \frac{1}{(2\pi \sigma^2)^\frac{3}{4}}~e^{\frac{-(\bold{x}-\bold{\bar{x}})^2}{4 \sigma^2}}~ e^{-i{\bold{k}}.(\bold{x}-\bar{\bold{x}})}~,
\eea
where $\bold{x}$ denotes spatial coordinates and $\bold{\bar{x}}$ is the central location of the laser beam. In the normalization factor $(2\pi \sigma^2)^{-\frac{3}{4}}$, the parameter $\sigma$ represents the width of the Gaussian wavepacket, and $\bold{k}$ refers to the mean momentum of the photons within the wavepacket. The Fourier transform of the photon field in the parallel arms of Fig. \ref{ring}a is defined as follows
 \cite{Ishikawa:2005, Ishikawa:2013, Ishikawa:2014, Ishikawa:2018} 
\bea
\hat{A}_\mu (x)=\int d^3\bold{p}~\frac{1}{(2\pi)^{3/2}}\frac{1}{\sqrt{2 p^0}} \left(\frac{1}{2\pi \beta_\perp^2} \right)^\frac{1}{2}~ \left(\frac{1}{2\pi \beta_{\|}^2}\right)^{\frac{1}{4}}e^{\frac{- (\bold{p}-\bold{\bar{p}})_\perp^2}{4 \beta_\perp^2}}~e^{\frac{- (p_z-\bar{p}_z)^2}{4 \beta_\|^2}}~\sum_s \left[\hat{a}_s(\bold{p}) \epsilon_\mu^s(\bold{p}) e^{-ip.(x-\bar{x})}+\text{h.c.}\right]~,  \label{Gaussian photon}
\eea
where $\hat{a}_s$ denotes the annihilation operator with the polarization index $s$, while $p = (p^0, \bold{p})$ represents the four-momentum, and $\epsilon_\mu$ corresponds to the polarization vector. Also here, $\beta_\perp$ and $\beta_\|$ are defined as the inverse of their corresponding widths, $\sigma$. The index $\|$ refers to the $z$ direction, while the index $\perp$ refers to the $x$ and $y$ directions. In the remainder of this section, we apply the wavepacket basis in addition to the plane-wave basis in our QFT calculations and compare the results. Using the wavepacket basis, the final Hamiltonian of photon-photon interaction mediated by a graviton follows that
\bea
\hat{H}_I(t)&=&\frac{ -4\kappa^2 }{(16 \pi )^2} \int d^4 x' \int d^3\bold{x} \int \frac{d^3\bold{p}}{(2\pi)^{3/2}} \frac{1}{\sqrt{2 p^0}} \int \frac{d^3\bold{p'}}{(2\pi)^{3/2}}  \frac{1}{\sqrt{2 {p'}^0}}
 \int \frac{d^3\bold{q}}{(2\pi)^{3/2}} \frac{1}{\sqrt{2 q^0}} \int \frac{d^3\bold{q'}}{(2\pi)^{3/2}} \frac{1}{\sqrt{2 {q'}^0}}\int \frac{d^4k}{(2\pi)^4} \nonumber\\
&\times &\left(\frac{1}{2\pi \beta_\perp^2}\right)^2 \left(\frac{1}{2\pi \beta_{\|}^2}\right)~\sum_{s , s' =1}^{2} \sum_{r,r' =1}^{2} e^{i(p'-p).(x-\bar{x})} ~ e^{-i(q-q').(x'-\bar{x}')} \frac{i e^{-ik.(x-x')}}{k^2}  \,~\hat{a}^\dagger_{s'}(p') \hat{a}_{s}(p) \hat{a}^\dagger_{r'}(q') \hat{a}_{r}(q)\nonumber\\
&\times& e^{\frac{-(\bold{q'}-\bold{\bar{q}})_\perp^2}{4 \beta_\perp^2}}  e^{\frac{-(\bold{p'}-\bold{\bar{p}})_\perp^2}{4 \beta_\perp^2}} e^{\frac{-(\bold{q}-\bold{\bar{q}})_\perp^2}{4 \beta_\perp^2}} e^{\frac{-(\bold{p}-\bold{\bar{p}})_\perp^2}{4 \beta_\perp^2}}~e^{\frac{-(q_z'-\bar{q}_z)^2}{4 \beta_\|^2}}  e^{\frac{-(p'_z-\bar{p}_z)^2}{4 \beta_\|^2}} e^{\frac{-(q_z-\bar{q}_z)^2}{4 \beta_\|^2}}  e^{\frac{-(p_z-\bar{p}_z)^2}{4 \beta_\|^2}} \nonumber\\
&\times& \{(p.q) (q'.p') \left[(\epsilon_s(p).\epsilon_{s'}(p')) (\epsilon_{r'}(q').\epsilon_r(q))+(\epsilon_s(p).\epsilon_{r'}(q')) (\epsilon_r(q).\epsilon_{s'}(p')) \right] \nonumber\\
&&+ (\epsilon_s(p).\epsilon_r(q)) (\epsilon_{s'}(p').\epsilon_{r'}(q'))] [(p.p')(q.q')+(p.q')(q.p')-4(p.q)(p'.q')]\}~. \label{Hamiltoni1}
\eea
Assuming that photons around the ring maintain a fixed circular polarization (indicated by \cite{Zain:2023}) the circular polarization vectors for the right arm of the ring are given as follows
\bea
\epsilon_R(p)=\frac{1}{\sqrt{2}}\left(
                   \begin{array}{cc}
                   1 ~\\
                  i  ~ \\
                  \end{array}
                 \right)\, , \,\,\,\,\,~~
\epsilon_L(p)=\frac{1}{\sqrt{2}}\left(
                   \begin{array}{cc}
                   1 ~\\
                  -i  ~ \\
                 \end{array}
                 \right)\, ,
\eea
and also for the left arm of the ring as
\bea
\epsilon_R(q)=\frac{1}{\sqrt{2}}\left(
                   \begin{array}{cc}
                   -1 ~\\
                  i  ~ \\
                \end{array}
                 \right)\, , \,\,\,\,\,~~\epsilon_L(q)=\frac{1}{\sqrt{2}}\left(
                   \begin{array}{cc}
                   -1 ~\\
                  -i  ~ \\
                \end{array}
                 \right)\, ,
\eea 
where $L$ and $R$ denote left- and right-handed circular polarizations, respectively. Therefore the only non-vanishing contribution in Eq. \eqref{Hamiltoni1} comes from the first term $\left( \epsilon_s(p)\cdot\epsilon_{s'}(p')\right) \left( \epsilon_{r'}(q')\cdot\epsilon_r(q)\right)=1$, while the others vanish.
For instance, $ \epsilon_L(p)\cdot\epsilon_{L}(q') = \epsilon_{R}(p)\cdot\epsilon_R(q')=0$.
Therefore, assuming $p^0\simeq|p_z|$ \footnote{The energy of a photon is given by $p^0=\sqrt{p_x^2+p_y^2+p_z^2}$. Assuming $p_z\gg p_x \sim p_y$, this simplifies to $p^0\simeq|p_z|$.}, the final Hamiltonian is expressed as follows (see Appendix \ref{graviton} for a detailed derivation)
\bea
\hat{H}_I(t)&\simeq&\frac{ -4G }{{p_0}^2 L^2 \sigma^4} \left[ \sigma^{4} {{p_0}}^4 \mathcal{A}_0(\sigma)+2\sigma^{2}{p_0 }^2 \mathcal{A}_1(\sigma)+4\mathcal{A}_3(\sigma)\right]  \sum_{s,r=1}^2  \hat{a}^\dagger_{s}({p_0})~ \hat{a}_{s}({p_0}) ~\hat{a}^\dagger_{r}(-{p_0}) ~\hat{a}_{r}(-{p_0})~.  \label{GHamilton1}
\eea
Numerical investigations indicate that, in the regime where $\sigma / W \ll 1$, the coefficient $\mathcal{A}_0$ scales as $\mathcal{A}_0 \sim 2L\ln(L/W)$ while $\mathcal{A}_1$ and $\mathcal{A}_3$ scale as $\mathcal{A}_1, \mathcal{A}_3 \sim \sigma^4/W^3$. 
In the limit where $\sigma$ is sufficiently small, the first term of Eq. \eqref{GHamilton1} reduces to the plane-wave Hamiltonian (1), while the second term vanishes. Furthermore, in the regime where ${p_0}^4 W^3 L\gg1$, the third term which contains $\mathcal{A}_3$, becomes negligible compared to the first term, which contains $\mathcal{A}_0$. Consequently, in these regimes, only the first term remains significant, and the plane-wave results are recovered. Hereafter, we refer to the second and third terms of Eq. \eqref{GHamilton1} as the wavepacket correction to the Hamiltonian (1).
\subsection{Minimum required laser power}
We now determine the minimum required laser power when the corrections due to considering a wavepacket form are included. As discussed in \cite{Zain:2023}, the required laser power, given by ${P} \approx N \hbar \omega_0 c / 2L$, to achieve quantum parameter estimation of $\chi_{\text{Q}}$ is
\bea
{P} \gtrsim \frac{c^3}{16}\left( \frac{c \hbar ^2}{12 L^3 T \mathcal{F} G^2 \left[\ln(L/W)\right]^2}\right)^{1/4},\label{Power-gravity}
\eea
where $T$ is the total time of experiment. For $L=10\, \text{km}$, $W=10\, \text{cm}$, $\mathcal{F}=450$ and $T$ about one year, this yields
\bea
{P} \gtrsim 125 ~\text{MW}~.
\eea
In the presence of a Gaussian consideration, the Hamiltonian, in SI units, is
\bea
\hat{H}_I&\simeq&- 8 G \left[\left(\frac{\hbar \omega_0}{L c^2}\right)^2  \mathcal{A}_0(\sigma)+2\left(\frac{ \hbar}{\sigma L c}\right)^2  \mathcal{A}_1(\sigma)+4\left(\frac{ \hbar}{\sigma^2 L \omega_0}\right)^2  \mathcal{A}_3(\sigma)\right]~\hat{a}^\dagger \hat{a}\hat{a}^\dagger  \hat{a}~.
\eea
The laser power calculation using this Hamiltonian, with parameters $\sigma=1\,\text{mm}$ and a laser frequency of $\nu_0 \simeq 150 ~\text{THz}$, yields
\bea
{P} \gtrsim \frac{c^2 }{16}\left( \frac{ c \hbar ^2}{12  L^3 T \mathcal{F} G^2 \left[\frac{\mathcal{A}_0}{2 Lc^2}+\frac{2\mathcal{A}_1(\sigma)}{L\sigma^2 \omega_0^2}+\frac{4 c^2 \mathcal{A}_3(\sigma)}{L\sigma^4 \omega_0^4}\right]^2} \right)^{1/4}\simeq 125~\text{MW}~.
\eea
Therefore, clearly the wavepacket corrections are negligible in the results of photon self-interaction mediated by gravitons. Nevertheless, as will become clear shortly, interactions arising from other effects (e.g., interactions with axions) could surpass those related to gravitons.

\section{Axionic self-interaction of photons}\label{S-AXION}

\begin{figure}
   \includegraphics[width=1.8in]{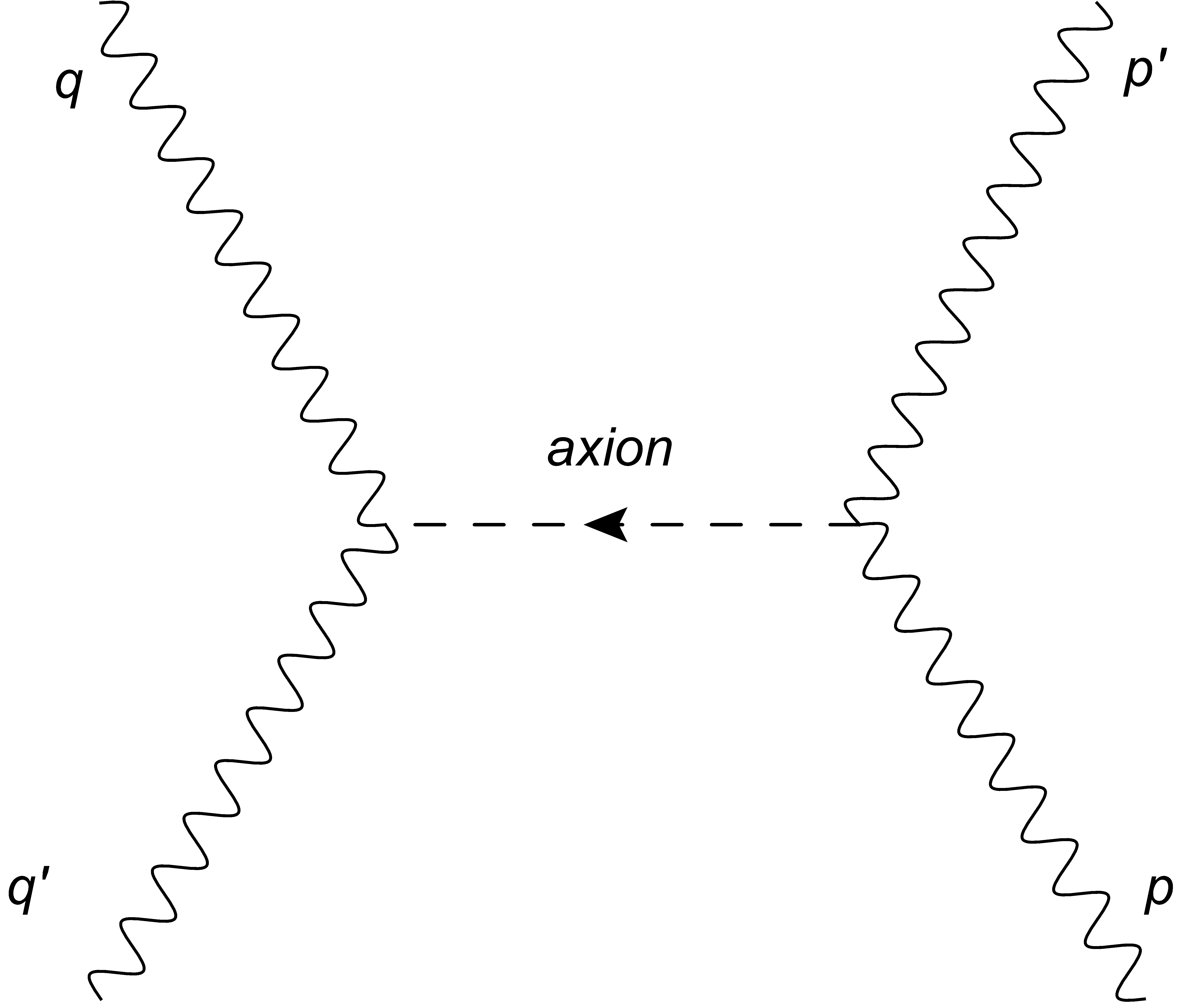}\\
  \caption{Feynmann diagram for the general $t$-channel photon-photon interaction mediated by the axion field. }\label{feyndiag}
\end{figure}

Now, we consider the effect of wavepacket corrections on photon self-interaction mediated by the axion field.  
To begin, we first extract the relevant photon-photon Hamiltonian for a general case in the \( t \)-channel Feynman diagram, and then proceed to our specialized setup. To do this, one starts with the Lagrangian for the photon-axion vertex in natural units \cite{Raffelt:1988}
\bea
\mathcal{L}_{a \gamma \gamma}=-\frac{1}{4} g_{a\gamma \gamma} ~ a~F_{\mu \nu}(x) \widetilde{F}_{\mu \nu}(x)~, \label{ax-ph vertex}
\eea
where $a$ is the axion field, $g_{a\gamma \gamma}$ is the coupling constant of the photon-axion interaction and $\widetilde{F}_{\alpha \beta}(x) = \frac{1}{2} \varepsilon^{\mu \nu \alpha \beta} {F}_{\alpha \beta}(x)$, where $\varepsilon^{\mu \nu \alpha \beta}$ is the LeviCivita tensor. By combining these vertices as building blocks, various Feynman diagrams can be created, each relevant to specific physical phenomena. In this paper, we focus on the interaction between photons through the exchange of virtual axion particles.
For a general photon-photon interaction mediated by an axion field, shown in Fig. \ref{feyndiag}, we have the following relation for the corresponding second-order $S$-matrix
\bea
\mathcal{S}^{(2)}=-\frac{1}{2}\int_{-\infty}^{+ \infty}dt \int_{-\infty}^{+ \infty}dt'~ \text{T}\{H_{a\gamma \gamma}(t) H_{a \gamma \gamma} (t')\} = -i \int_{-\infty}^{+ \infty}dt~ H^{(2)}(t)~.
\eea
Therefore, the final corresponding Hamiltonian for this setup is
\bea
H_I(t)=\frac{ g^2_{a\gamma \gamma} }{16} \int d^4 x' \int d^3\bold{x}~  F_{\mu \nu}(x) ~\widetilde{F}_{\mu \nu}(x) D_a(x-x') F_{\rho \sigma}(x') \widetilde{F}_{\rho \sigma}(x')~, \label{ax-ph tree level}
\eea
where

\bea
 D_a(x-x')=\int \frac{d^4k}{(2\pi)^4}~\frac{i e^{-ik.(x-x')}}{k^2-m_a^2+ik^0\Gamma_\beta(k^0)}~,
\eea
is the axion propagator, where $m_a$ denotes the axion mass and $\Gamma_\beta$ the axion decay rate.
Now, in the plane-wave basis and after some calculations, for a general photon direction (see Fig. \ref{feyndiag}), we obtain
\bea
\hat{H}_{I}(t)&=&\frac{ g^2_{a\gamma \gamma} }{4} \int d^4 x' \int d^3\bold{x} \int \frac{d^3\bold{p}}{(2\pi)^{3/2}}\frac{1}{\sqrt{2p^0}} \int \frac{d^3\bold{p'}}{(2\pi)^{3/2}} \frac{1}{\sqrt{2p'^0}} \int \frac{d^3\bold{q}}{(2\pi)^{3/2}}\frac{1}{\sqrt{2q^0}} \int \frac{d^3\bold{q'}}{(2\pi)^{3/2}}\frac{1}{\sqrt{2q'^0}}\int \frac{d^4k}{(2\pi)^4} \nonumber\\
&\times &\sum_{s , s' =1}^{2} \sum_{r,r' =1}^{2} e^{i(p'-p).x} e^{-i(q-q').x'}  \frac{i e^{-ik.(x-x')}}{k^2-m_a^2+ik^0\Gamma_\beta(k^0)}~ \hat{a}^\dagger_{s'}(p') \hat{a}_{s}(p) \hat{a}^\dagger_{r'}(q') \hat{a}_{r}(q)\nonumber\\
&\times&  \left\{(p.q')\left\{(p'.\epsilon_r)[(q.\epsilon_{s'}) (\epsilon_s.\epsilon_{r'})-(q.\epsilon_s)(\epsilon_{s'}.\epsilon_{r'})]+(p'.q)[(\epsilon_r.\epsilon_{s}) (\epsilon_{s'}.\epsilon_{r'})-(\epsilon_r.\epsilon_{s'})(\epsilon_{r'}.\epsilon_{s})] \right. \right.\nonumber\\
&+& \left. \left. (p'.\epsilon_{r'})[(q.\epsilon_{s}) (\epsilon_{s'}.\epsilon_r)-(q.\epsilon_{s'})(\epsilon_{s}.\epsilon_{r})]\right\} -(p.\epsilon_r)(\epsilon_{r'}.\epsilon_{s'})(q.p')(q'.\epsilon_s)\right.\nonumber\\
&+& \left.(p.q)(p'.\epsilon_r)(\epsilon_{r'}.\epsilon_{s'})(q'.\epsilon_s)+(p.\epsilon_r)(q.\epsilon_{s'})(p'.\epsilon_{r'})(q'.\epsilon_s)-(p.q)(p'.\epsilon_{r'})(q'.\epsilon_s)(\epsilon_r. \epsilon_{s'})\right.\nonumber\\
&+& \left.(p.\epsilon_r)(q.p')(q'.\epsilon_{s'})(\epsilon_{r'}.\epsilon_s)-(p.q)(p'.\epsilon_r)(q'.\epsilon_{s'})(\epsilon_s.\epsilon_{r'})-(p.\epsilon_r)(q.\epsilon_s)(p'.\epsilon_{r'})(q'.\epsilon_{s'})\right.\nonumber\\
&+& \left.(p.q)(p'.\epsilon_{r'})(q'.\epsilon_{s'})(\epsilon_s.\epsilon_r)+(p'.q')\left\{(p.\epsilon_r)[(q.\epsilon_{s}) (\epsilon_{s'}.\epsilon_{r'})-(q.\epsilon_{s'})(\epsilon_{s}.\epsilon_{r'})] \right.\right.\nonumber\\
&+& \left.\left.(q.p)[(\epsilon_r.\epsilon_{s'}) (\epsilon_{s}.\epsilon_{r'})-(\epsilon_r.\epsilon_{s})(\epsilon_{s'}.\epsilon_{r'})]\right\}+(p.\epsilon_{r'})[(p'.\epsilon_r)[(q.\epsilon_{s}) (q'.\epsilon_{s'})-(q.\epsilon_{s'})(q'.\epsilon_s)]\right.\nonumber\\
&+& \left.(p'.q)[(\epsilon_r.\epsilon_{s'}) (q'.\epsilon_{s})-(\epsilon_r.\epsilon_{s})(q'.\epsilon_{s'})]+ (p'.q')[(q.\epsilon_{s'}) (\epsilon_{s}.\epsilon_r)-(q.\epsilon_{s})(\epsilon_{s'}.\epsilon_{r})]\right\}~.  \label{a-p-general}
\eea 
For the momentum directions corresponding to the ring cavity, illustrated in Fig. \ref{ring}b, the expression \eqref{a-p-general} simplifies as follows
\bea
\hat{H}_I(t)&=&\frac{ g^2_{a\gamma \gamma} }{4} \int d^4 x' \int d^3\bold{x} \int \frac{d^3\bold{p}}{(2\pi)^{3/2}}\frac{1}{\sqrt{2p^0}} \int \frac{d^3\bold{p'}}{(2\pi)^{3/2}} \frac{1}{\sqrt{2p'^0}} \int \frac{d^3\bold{q}}{(2\pi)^{3/2}}\frac{1}{\sqrt{2q^0}} \int \frac{d^3\bold{q'}}{(2\pi)^{3/2}}\frac{1}{\sqrt{2q'^0}}\int \frac{d^4k}{(2\pi)^4} \nonumber\\
&\times &\sum_{s , s' =1}^{2} \sum_{r,r' =1}^{2} e^{i(p'-p).x} e^{-i(q-q').x'}  \frac{i e^{-ik.(x-x')}}{k^2-m_a^2+ik^0\Gamma_\beta(k^0)}~ \hat{a}^\dagger_{s'}(p') \hat{a}_{s}(p) \hat{a}^\dagger_{r'}(q') \hat{a}_{r}(q)\nonumber\\
&\times& \{[(p.q') (q.p')- (p.q) (p'.q')]  [(\epsilon_r(q).\epsilon_s(p)) (\epsilon_r'(q').\epsilon_s'(p')) - (\epsilon_r(q).\epsilon_s'(p')) (\epsilon_r'(q').\epsilon_s(p))]\}~. \label{axi-Hamiltoni}
\eea

Under the forward scattering condition $q_z= q'_z$ and $p_z = p'_z$, this Hamiltonian vanishes because the factor $
\big[(p \cdot q')(q \cdot p') - (p \cdot q)(p' \cdot q')\big]
$ equals zero. Therefore, the chosen mechanism and configuration for axion detection are ineffective in the forward scattering case.\\ Nonetheless, an important question arises: Does the Hamiltonian remain null when a wavepacket photon beam is used instead of plane-wave photons? We address this question in the following subsection. Before proceeding, it is crucial to note that the axion-photon vertex takes the form $\epsilon^{\mu\nu\alpha\beta} p_\mu p'_\nu \epsilon_{r\alpha} \epsilon_{r'\beta}$.
This expression vanishes in an exact plane-wave beam where $p = p'$. However, in a beam with finite width, the incoming and outgoing momenta may differ slightly, potentially leading to a non-zero amplitude.\\

\subsection{Axionic self-interaction of photons in wavepacket basis}

According to Fig.~\ref{ring}a, with the axion field as a mediator and the photon field is represented as a Gaussian wavepacket \eqref{Gaussian photon} (see Appendix \ref{axionapp}), the dominant term yields
\bea
\hat{H}_I (t)&\simeq&\frac{ g^2_{a\gamma \gamma} }{16} \int \frac{dp_z}{(2\pi)^{3/2}} \int \frac{dp'_z}{(2\pi)^{3/2}}~ \int \frac{dq_z}{(2\pi)^{3/2}}\int \frac{dq'_z}{(2\pi)^{3/2}}~\int_{-\infty}^\infty dz ~\int_{-\infty}^\infty dz' ~ \frac{e^{i(p'^{0}-p^0+q^0-q'^0)t}}{\sqrt{p^0 q^0 p'^0 q'^0}} \sum_{s , s' =1}^{2} \sum_{r,r' =1}^{2} \nonumber\\
&\times&  \left(\frac{1}{2\pi \beta_\perp^2}\right)^2 \left(16\pi^2 \beta_\perp^6\right)^2~\int d{\bold{x_\perp}^2} \int d{\bold{x'}^{2}_\perp}~e^{-2\beta_\perp^2[(x-W)^2+y^2]}~e^{-2\beta_\perp^2[x'^2+y'^2]}~\frac{e^{-m_a\sqrt{(\bold{x}_\perp - \bold{x_\perp'})^2+ (z-z')^2}}}{\sqrt{(\bold{x}_\perp - \bold{x_\perp'})^2+ (z-z')^2}}\nonumber\\
&\times& \left(\frac{1}{2\pi \beta_{\|}^2}\right)
~e^{\frac{-(q_z'-\bar{q}_z)^2}{4 \beta_\|^2}}  e^{\frac{-(p'_z-\bar{p}_z)^2}{4 \beta_\|^2}} e^{\frac{-(q_z-\bar{q}_z)^2}{4 \beta_\|^2}}  e^{\frac{-(p_z-\bar{p}_z)^2}{4 \beta_\|^2}}~e^{i(p'_z-p_z).(z-\bar{z})} ~e^{-i(q_z-q'_z).(z'-\bar{z}')}\nonumber\\
& \times&  \hat{a}^\dagger_{s'}(p') \hat{a}_{s}(p) \hat{a}^\dagger_{r'}(q') \hat{a}_{r}(q) \{[(\epsilon_r(q).\epsilon_s(p)) (\epsilon_{r'}(q').\epsilon_{s'}(p')) - (\epsilon_r(q).\epsilon_{s'}(p')) (\epsilon_{r'}(q').\epsilon_s(p))]\}\nonumber\\
&\times&\left\{4 x' (x-W)(q'^0 p^0+q'_z p_z+q^0 p'^0+q_zp'_z+q^0 p^0+q_z p_z+p'^0 q'^0+p'_z q'_z-4 \bar{p}_x^2)+16\bar{p}_x^2 x' y\right. \nonumber\\ 
&+&\left.4 y'\left[y(q'^0 p^0+q'_z p_z+q^0 p'^0+q_zp'_z+q^0 p^0+q_z p_z+p'^0 q'^0+p'_z q'_z+4 \bar{p}_x^2)+4\bar{p}_x^2 (W-x)\right] \right. \nonumber\\
&+&\left. \frac{2 i}{\beta^2}\left[-(q^0 p^0+q_z p_z-p'^0 q'^0-p'_z q'_z) \left[\bar{p}_x(W-x-x'+y-y') \right]\right. \right. \nonumber\\
&+& \left. \left. (q'^0 p^0+q'_z p_z-q^0 p'^0-q_zp'_z)\left[\bar{p}_x (W-x+x'+y+y') \right]\right]  \right. \nonumber\\
 &+&  \left. \frac{1}{\beta^4} \left[-(p^0q^0+p_z q_z) (p'^0 q'^0+p'_z q'_z)
+(q'^0p^0+q'_z p_z)(q^0 p'^0+q_z p'_z)\right] \right\}~. \label{axion-hamiltonian}
\eea
Here, we have ignored the axion decay rate $\Gamma_\beta$, which is trivial in the propagator denominator. Assuming the Gaussian distribution can be expressed as a discrete summation of Dirac delta functions (see Appendix \ref{axionapp}), we have
\bea
\hat{H}_I (t)&\simeq&\frac{ g^2_{a\gamma \gamma} }{16} \int \frac{dp_z}{(2\pi)^{3/2}} \int \frac{dp'_z}{(2\pi)^{3/2}} \int \frac{dq_z}{(2\pi)^{3/2}}\int \frac{dq'_z}{(2\pi)^{3/2}}\int_{-\infty}^\infty dz \int_{-\infty}^\infty dz' ~ \frac{e^{i(p'^{0}-p^0+q^0-q'^0)t}}{\sqrt{p^0 q^0 p'^0 q'^0}} \sum_{s , s' =1}^{2} \sum_{r,r' =1}^{2} \nonumber\\
&\times& \left(\frac{1}{2\pi \beta_{\|}^2}\right)
~e^{\frac{-(q_z'-\bar{q}_z)^2}{4 \beta_\|^2}}  e^{\frac{-(p'_z-\bar{p}_z)^2}{4 \beta_\|^2}} e^{\frac{-(q_z-\bar{q}_z)^2}{4 \beta_\|^2}}  e^{\frac{-(p_z-\bar{p}_z)^2}{4 \beta_\|^2}}~e^{i(p'_z-p_z).(z-\bar{z})} ~e^{-i(q_z-q'_z).(z'-\bar{z}')}\nonumber\\
&\times& \left(16\pi^4 \beta_\perp^4\right)~ \hat{a}^\dagger_{s'}(p') \hat{a}_{s}(p) \hat{a}^\dagger_{r'}(q') \hat{a}_{r}(q) \{[(\epsilon_r(q).\epsilon_s(p)) (\epsilon_{r'}(q').\epsilon_{s'}(p')) - (\epsilon_r(q).\epsilon_{s'}(p')) (\epsilon_{r'}(q').\epsilon_s(p))]\}\nonumber\\
&\times&~\frac{4(q'^0 p^0+|q'_z| |p_z|+q^0 p'^0+|q_z| |p'_z|+q^0 p^0+|q_z| |p_z|+p'^0 q'^0+|p'_z| |q'_z|)}{\beta_\perp^2}\left[\frac{ \mathcal{P}_1~ e^{-m_a \sqrt{W^2+ (z-z')^2}}}{\sqrt{W^2+ (z-z')^2}}\right.\nonumber\\
&+&\left. \frac{\mathcal{P}_2~ e^{-m_a \sqrt{(W-\sigma)^2+ (z-z')^2}}}{\sqrt{(W-\sigma)^2+ (z-z')^2}} +\frac{\mathcal{P}_2~ e^{-m_a \sqrt{(W+\sigma)^2+ (z-z')^2}}}{\sqrt{(W+\sigma)^2+ (z-z')^2}}-\frac{\mathcal{P}_3~ e^{-m_a \sqrt{(W-2\sigma)^2+ (z-z')^2}}}{\sqrt{(W-2\sigma)^2+ (z-z')^2}}+ \cdot \cdot \cdot\right]~, \nonumber\\
\eea
where the $\mathcal{P}_n$ coefficients come from the numerical calculations presented in Appendix \ref{axionapp}. Integrating over the momenta in the $z$ direction and assuming a single mode cavity with $|\bar{p}_z|=|\bar{q}_z|={p_0}$, then yields
\bea
\hat{H}_I (t)&\simeq&\frac{ g^2_{a\gamma \gamma} }{16} \frac{8 \pi^2}{L^2}\frac{32{p_0}^2}{\beta_\perp^2} \left(16\pi^4 \beta_\perp^4\right)~\int_{0}^L dz ~\int_{0}^L dz' ~ \frac{1}{{p_0}^2} \sum_{s , s' =1}^{2} \sum_{r,r' =1}^{2} \hat{a}^\dagger_{s'}(p_0) \hat{a}_{s}(p_0) \hat{a}^\dagger_{r'}(-p_0) \hat{a}_{r}(-p_0)  \nonumber\\
&\times&\{[(\epsilon_r(-p_0).\epsilon_s(p_0)) (\epsilon_{r'}(-p_0).\epsilon_{s'}(p_0)) - (\epsilon_r(-p_0).\epsilon_{s'}(p_0)) (\epsilon_{r'}(-p_0).\epsilon_s(p_0))]\} \left[\frac{ \mathcal{P}_1~ e^{-m_a \sqrt{W^2+ (z-z')^2}}}{\sqrt{W^2+ (z-z')^2}}\right.\nonumber\\
&+&\left. \frac{\mathcal{P}_2~ e^{-m_a \sqrt{(W-\sigma)^2+ (z-z')^2}}}{\sqrt{(W-\sigma)^2+ (z-z')^2}} +\frac{\mathcal{P}_2~ e^{-m_a \sqrt{(W+\sigma)^2+ (z-z')^2}}}{\sqrt{(W+\sigma)^2+ (z-z')^2}}-\frac{\mathcal{P}_3~ e^{-m_a \sqrt{(W-2\sigma)^2+ (z-z')^2}}}{\sqrt{(W-2\sigma)^2+ (z-z')^2}}+ \cdot \cdot \cdot\right]~. \nonumber\\ \label{AHamilton2}
\eea
Assuming a fixed photon polarization during the interaction, the Hamiltonian vanishes because the factor of
\bea
 [(\epsilon_r(-p_0).\epsilon_s(p_0)) (\epsilon_{r'}(-p_0).\epsilon_{s'}(p_0)) - (\epsilon_r(-p_0).\epsilon_{s'}(p_0)) (\epsilon_{r'}(-p_0).\epsilon_s(p_0))]~,\nonumber
\eea
cancels out in this case. Therefore, this interaction occurs only if the polarization of the photons changes as a result of the interaction. For simplicity, we consider one possible scenario: a Faraday rotation of approximately $90^\circ$ of the polarization plane, such as when $r = 1, r' = 2$, or vice versa (and similarly for $s, s'$). Therefore \eqref{AHamilton2} yields
\bea
\hat{H}_I (t)&\simeq&\frac{256 \pi^6 g^2_{a\gamma \gamma} }{L^2 \sigma^2} ~\sum_{s , s' =1}^{2} \sum_{r,r' =1}^{2} \hat{a}^\dagger_{s'} \hat{a}_{s} \hat{a}^\dagger_{r'} \hat{a}_{r} ~\int_{0}^L dz ~\int_{0}^L dz' ~ 
\left[\frac{\mathcal{P}_1~ e^{-m_a \sqrt{W^2+ (z-z')^2}}}{\sqrt{W^2+ (z-z')^2}}\right.\nonumber\\
&+&\left. \frac{\mathcal{P}_2~ e^{-m_a \sqrt{(W-\sigma)^2+ (z-z')^2}}}{\sqrt{(W-\sigma)^2+ (z-z')^2}} +\frac{\mathcal{P}_2~ e^{-m_a \sqrt{(W+\sigma)^2+ (z-z')^2}}}{\sqrt{(W+\sigma)^2+ (z-z')^2}}-\frac{\mathcal{P}_3~ e^{-m_a \sqrt{(W-2\sigma)^2+ (z-z')^2}}}{\sqrt{(W-2\sigma)^2+ (z-z')^2}}+ \cdot \cdot \cdot \right]~, \nonumber\\ 
\eea
which, in SI units, becomes 
\bea
\hat{H}_I (t)&\simeq&\frac{256 \pi^6 g'^2_{a\gamma \gamma} \hbar c^3}{ L^2 \sigma^2}~ \sum_{s , s' =1}^{2} \sum_{r,r' =1}^{2} \hat{a}^\dagger_{s'} \hat{a}_{s} \hat{a}^\dagger_{r'} \hat{a}_{r}~\int_{0}^L dz ~\int_{0}^L dz' ~ 
\left[\frac{ \mathcal{P}_1~ e^{-m'_a \sqrt{W^2+ (z-z')^2}}}{\sqrt{W^2+ (z-z')^2}}\right.\nonumber\\
&+&\left. \frac{\mathcal{P}_2~ e^{-m'_a \sqrt{(W-\sigma)^2+ (z-z')^2}}}{\sqrt{(W-\sigma)^2+ (z-z')^2}} +\frac{\mathcal{P}_2~ e^{-m'_a \sqrt{(W+\sigma)^2+ (z-z')^2}}}{\sqrt{(W+\sigma)^2+ (z-z')^2}}-\frac{\mathcal{P}_3~ e^{-m'_a \sqrt{(W-2\sigma)^2+ (z-z')^2}}}{\sqrt{(W-2\sigma)^2+ (z-z')^2}}+ \cdot \cdot \cdot \right]~, \nonumber\\  \label{AHamilton}
\eea
where in summations, $s\neq s'$ and $r\neq r'$, also $g'_{a\gamma \gamma}={\hbar g_{a\gamma \gamma}}/({1.6\times 10^{-10}})$ has units of time, while $g_{a\gamma \gamma}$ is expressed in $\text{GeV}^{-1}$, and $m'_a={1.6\times 10^{-19}\,m_a}/{c^2} $ with $m_a$ expressed in eV.

After investigating photon-photon interaction, one can calculate the time evolution of the photon density matrix to examine detection properties, such as changes in photon polarization and other effects. Studying the changes in polarization will be beneficial for designing the detection method using this setup in future research.

\subsection{Laser power requirement for Cramér-Rao condition}

As in the previous discussion regarding the graviton mediator, we now determine the minimum required laser power in the presence of a Gaussian correction term in the Hamiltonian for the axion mediator case, as shown in Eq. \eqref{AHamilton}. Therefore, we have
\bea
\hat{H}_I&\simeq&\frac{1024 \pi^6 g'^2_{a\gamma \gamma} \hbar c^3}{L^2 \sigma^2}~ \mathcal{B}(\sigma)~\hat{a}^\dagger \hat{a}\hat{a}^\dagger  \hat{a}~,
\eea
where
\bea
 \mathcal{B}(\sigma)=\int_0 ^L dz \int_0 ^L dz' \left[\frac{ \mathcal{P}_1~ e^{-m'_a \sqrt{W^2+ (z-z')^2}}}{\sqrt{W^2+ (z-z')^2}}+ \frac{\mathcal{P}_2~ e^{-m'_a \sqrt{(W-\sigma)^2+ (z-z')^2}}}{\sqrt{(W-\sigma)^2+ (z-z')^2}}+
\frac{\mathcal{P}_2~ e^{-m'_a \sqrt{(W+\sigma)^2+ (z-z')^2}}}{\sqrt{(W+\sigma)^2+ (z-z')^2}}+\cdot \cdot \cdot \right]~.\nonumber\\ \label{B}
\eea
Note that we can approximate the behavior of $\mathcal{B}$ in the regime of $\sigma / W \ll 1$ as $\mathcal{B} \sim \sigma^4/W^3$. Now, having the Fisher information and using the Cramér-Rao bound, we can obtain the variance of the desired parameters. One can follow the Fisher information calculation in Appendix \ref{appendix-QFI}. 

For $L=10~\text{km}, W=10~\text{cm}, \sigma=1~\text{mm}$, $\lambda=2~\mu\text{m}$, $\nu_0 \simeq 150 ~\text{THz}$, $\mathcal{F}=450$, in the range of the CAST experiment (e.g. $m_a=10^{-7} ~\text{eV},\, g_{a\gamma \gamma}=10^{-10}~\text{GeV}^{-1}$) and for a one-year total time period, the required laser power yields
\bea
{P}_{\chi_a} \gtrsim \frac{\hbar \omega_0 \sigma}{256 \pi^3 g'_{a\gamma\gamma}}\left( \frac{1}{3  L T c  \mathcal{F}  \mathcal{B}(\sigma)^2} \right)^{1/4} \simeq 2~\text{kW}~,
\eea
where ${P}_{\chi_a}$ denotes the power extracted from QFI of the $\chi_a$ parameter. Therefore, for a laser frequency of approximately 150 THz and a variance in the laser beam position within the tube of about $\sigma \sim 1~\text{mm}$, the minimum required laser power for both graviton and axion mediators is
\bea
{P}_{\text{grav}} \gtrsim 125 ~\text{MW}~,
\eea
\bea
{P}_{\text{axion}} \gtrsim 2 ~\text{kW}~.
\eea
Note that the power of 125 \,\text{MW} (reported in \cite{Zain:2023}), in the case of RC axion proposal, approximately would be achieved with $L\sim 40\, \text{cm}$, $W=5 \,\text{cm}$, $\sigma=1 \,\text{mm}$, $m_a\sim 10^{-7}\, \text{eV}$, $g_{a }\sim 10^{-10}\, \text{GeV}^{-1}$, $\mathcal{F}<100$, $\lambda=2\,\mu \text{m}$, $\nu_0\simeq150\, \text{THz}$, and during less than one minute of interrogation time.

\begin{figure}
   \includegraphics[width=7 in]{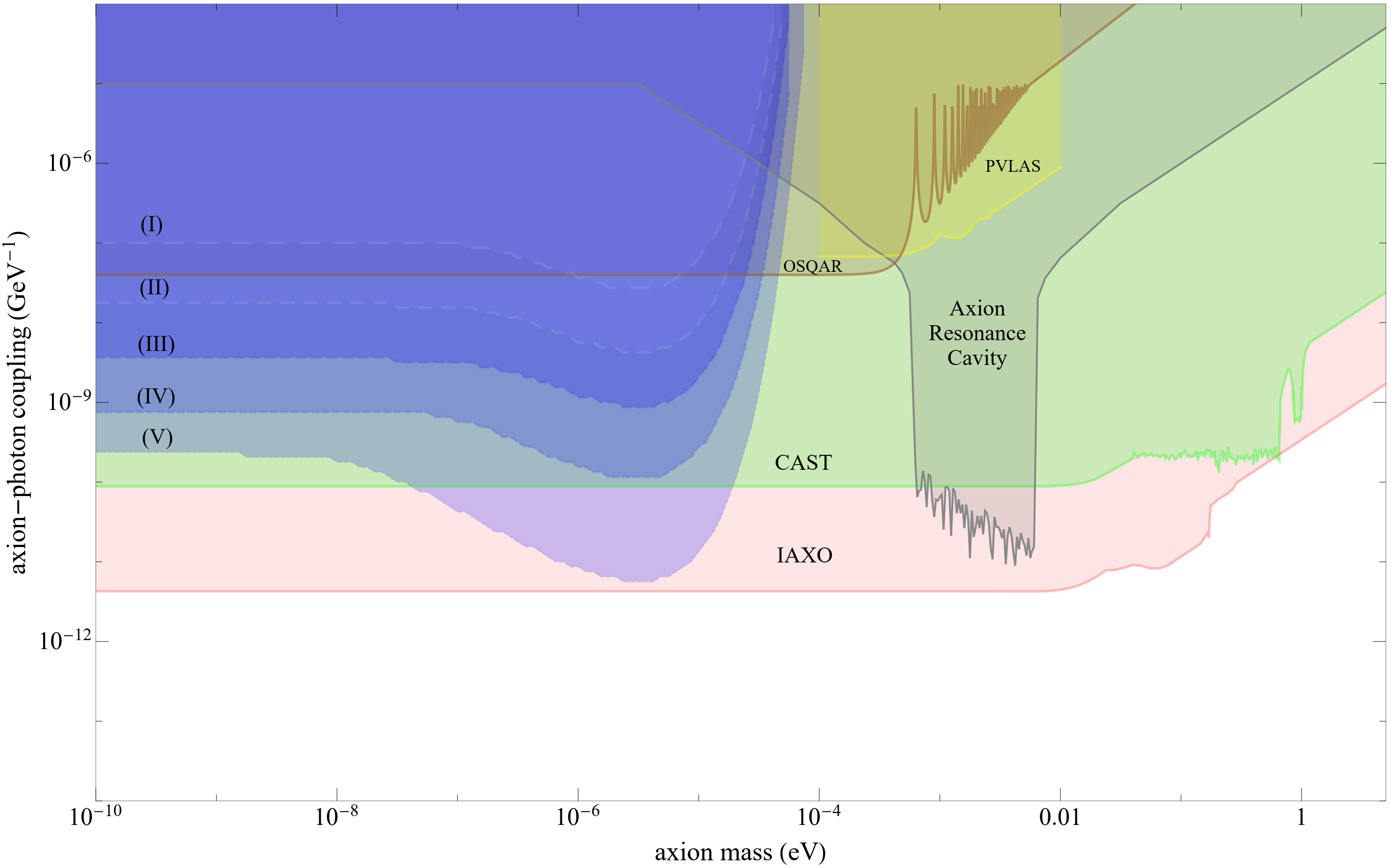}\\
  \caption{Exclusion regions in the plane of axion mass and axion-photon coupling constant for the CAST, IAXO, PVLAS, OSQAR-LSW, Axion Resonance Cavity \cite{Zarei:2022}, and the RC proposal (using the QFI of $\mathcal{F}_{\chi_a}$ \eqref{Fxa}) are shown. The area denoted as (I) corresponds to the RC proposal for ${P}=10\, \text{kW}$, $\mathcal{F}=300$, $L=3\, \text{m}$, $W=10\,\text{cm}$, $\sigma=1\, \text{mm}$, and $T\sim 3$ hours. Area (II) represents the RC proposal with ${P}=10\,\text{kW}$, $\mathcal{F}=300$, $L=3\, \text{m}$, $W=10\, \text{cm}$, and $T\sim 6$ months. Area (III) corresponds to the RC proposal with ${P}=300\,\text{kW}$, $\mathcal{F}=300$, $L=3\, \text{m}$, $W=10\, \text{cm}$, and $T\sim 3$ hours. Area (IV) relates to the RC proposal with ${P}=300\,\text{kW}$, $\mathcal{F}=300$, $L=10\, \text{m}$, $W=10\,\text{cm}$, and $T\sim 14$ days. Finally, area (V) represents the RC proposal with ${P}=500\,\text{kW}$, $\mathcal{F}=400$, $L=100\,\text{m}$, $W=10\, \text{cm}$, and $T\sim 6$ months.}\label{excl1} 
\end{figure}

From the Fisher information (Appendix \ref{appendix-QFI}), and considering the predicted parameter variance using the Cramér-Rao bound, we can obtain a bound on the axion mass and axion-photon coupling constant related to the results of previous sections for the RC proposal, as shown in Fig. \ref{excl1}. Apparently, the excluded limits of the $m_a-g_{a \gamma \gamma}$ region obtained from this realization proposal, in short interrogation time, are not as sensitive as the results of the CAST experiment. Although they are better than those provided, for example, by the LSW techniques of axion detection, they are still not optimal. Nevertheless, under conditions of high laser power, along with a long arm and extended interrogation time in the RC proposal, the CAST results are easily achieved. This RC proposal could be introduced as a candidate for a virtual axion detector. However, considering the exclusion limit obtained by the RC proposal, it would be more sensitive in the lower axion mass range, specifically below $10^{-4}$ eV. In other words, this RC proposal would be effective for low mass detections.

\section{Discussion and Conclusion}\label{S-CONCL}

To summarize, we have demonstrated that considering quantum interactions, whether with or without the wavepacket basis, leads to different results. This is evident in the case of axion exchange particles when comparing results from equations \eqref{axi-Hamiltoni} and \eqref{AHamilton}, although these changes are negligible for the graviton. The formalism of QFT uncovers precise details of interactions that may be obscured by other methods. Our study of the wavepacket assumption for beams provides insight into the need for greater accuracy in real quantum entity detectors, such as tabletop tests for quantum gravity. This improvement will enable the distinction and, if needed, suppression of effects related to wavepackets. Additionally, this research can be viewed as a technique for axion detection, particularly for low-mass axions. It is important to note that our proposal is most efficient for axion masses up to $10^{-4}$ eV.
\begin{acknowledgements}
A. H. acknowledges support from the Iran Science Elites Federation through a postdoctoral fellowship. Additionally, A. H. would like to thank M. Sharifian for the insightful discussions on numerical calculations. The work of M. A. was partially supported by the Yangyang Development Foundation.
\end{acknowledgements}
\begin{appendices}
\section{Calculation of the photon-photon Hamiltonian mediated by a graviton} \label{graviton}
\begin{figure}
   \includegraphics[width=4.5 in]{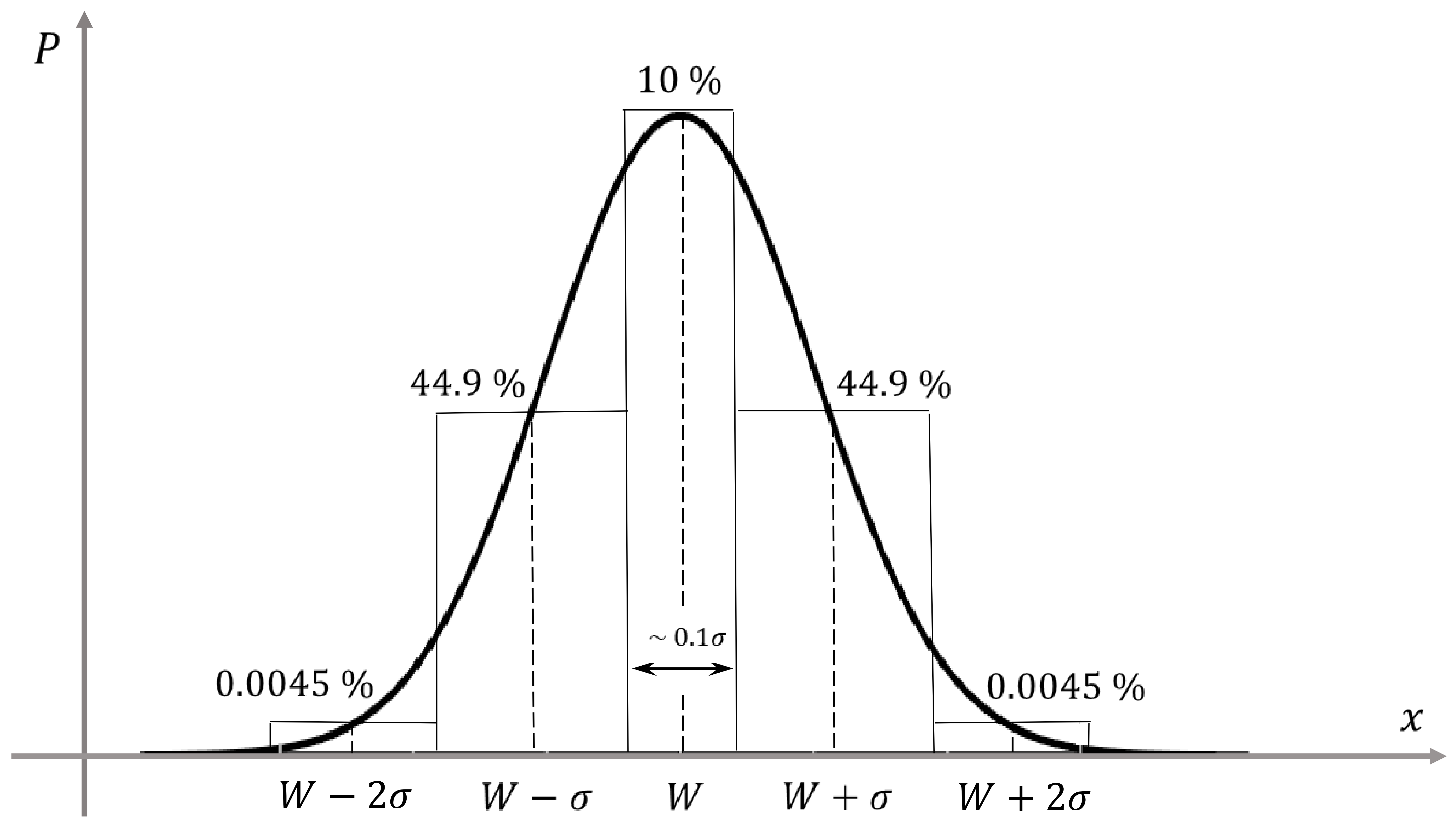}\\
  \caption{Gaussian probability distribution as described in Eq. \eqref{Gaussian-gravity} and its discrete approximation for a tube centerated at $x=W$. }\label{discrete}
\end{figure}

In this Appendix, we extract the Hamiltonian for the gravitational self-interaction of photons using the wavepacket expansion. Starting with the Hamiltonian \eqref{Hamiltoni1}
\bea
\hat{H}_I(t)&=&\frac{ -4\kappa^2 }{(16 \pi )^2} \int d^4 x' \int d^3\bold{x} \int \frac{d^3\bold{p}}{(2\pi)^{3/2}}\frac{1}{\sqrt{2 p^0}} \int \frac{d^3\bold{p'}}{(2\pi)^{3/2}} \frac{1}{\sqrt{2 {p'}^0}} \int \frac{d^3\bold{q}}{(2\pi)^{3/2}}\frac{1}{\sqrt{2 q^0}} \int \frac{d^3\bold{q'}}{(2\pi)^{3/2}} \frac{1}{\sqrt{2 {q'}^0}}\int \frac{d^4k}{(2\pi)^4} \nonumber\\
&\times &\left(\frac{1}{2\pi \beta_\perp^2}\right)^2 \left(\frac{1}{2\pi \beta_{\|}^2}\right)~\sum_{s , s' =1}^{2}~ \sum_{r,r' =1}^{2} e^{i(p'-p).(x-\bar{x})} e^{-i(q-q').(x'-\bar{x}')}~ \hat{a}^\dagger_{s'}(p') \hat{a}_{s}(p) \hat{a}^\dagger_{r'}(q') \hat{a}_{r}(q)\times [(p.q) (q'.p')] \nonumber\\
&\times&\frac{i e^{-ik.(x-x')}}{k^2} ~e^{\frac{-(\bold{q'}-\bold{\bar{q}})_\perp^2}{4 \beta_\perp^2}}  e^{\frac{-(\bold{p'}-\bold{\bar{p}})_\perp^2}{4 \beta_\perp^2}} e^{\frac{-(\bold{q}-\bold{\bar{q}})_\perp^2}{4 \beta_\perp^2}} e^{\frac{-(\bold{p}-\bold{\bar{p}})_\perp^2}{4 \beta_\perp^2}}~e^{\frac{-(q_z'-\bar{q}_z)^2}{4 \beta_\|^2}}  e^{\frac{-(p'_z-\bar{p}_z)^2}{4 \beta_\|^2}} e^{\frac{-(q_z-\bar{q}_z)^2}{4 \beta_\|^2}}  e^{\frac{-(p_z-\bar{p}_z)^2}{4 \beta_\|^2}}~, 
\eea
and after integrating over the variable of $t'$, with assuming the fixed circular polarization of the photons, we obtain
\bea
\hat{H}_I(t)&=&\frac{ -\kappa^2 }{(16 \pi )^2} \int d^3 \bold{x}' \int d^3 \bold{x} \int \frac{d^4k}{(2\pi)^4}  \int \frac{d^3\bold{p}}{(2\pi)^{3/2}}\int \frac{d^3\bold{p'}}{(2\pi)^{3/2}}
 \int \frac{d^3\bold{q}}{(2\pi)^{3/2}}\int \frac{d^3\bold{q'}}{(2\pi)^{3/2}}~ \frac{ e^{i(p'^{0}-p^0-k^0)t} }{\sqrt{p^0 q^0 p'^0 q'^0}}~ 2\pi \delta (q'^{0}-q^0+k^0)\nonumber\\
&\times &   \frac{e^{i\bold{k}.(\bold{x}-\bold{x}')}}{(k^{0})^2-\bold{k}^2}~ e^{-i(\bold{p'-p})_\perp \cdot(\bold{x}_\perp-\bold{\bar{x}}_\perp)}~ e^{-i(p'_z -p_z) (z-\bar{z})}~e^{-i(\bold{q'-q})_\perp \cdot (\bold{x'}_\perp-\bold{\bar{x}'}_\perp)}~ e^{-i(q'_z-q_z) (z'-\bar{z}')}  \nonumber\\
&\times&\left(\frac{1}{2\pi \beta_\perp^2}\right)^2 \left(\frac{1}{2\pi \beta_{\|}^2}\right)~e^{\frac{-(\bold{q'}-\bold{\bar{q}})_\perp^2}{4 \beta_\perp^2}}  e^{\frac{-(\bold{p'}-\bold{\bar{p}})_\perp^2}{4 \beta_\perp^2}} e^{\frac{-(\bold{q}-\bold{\bar{q}})_\perp^2}{4 \beta_\perp^2}} e^{\frac{-(\bold{p}-\bold{\bar{p}})_\perp^2}{4 \beta_\perp^2}}~e^{\frac{-(q_z'-\bar{q}_z)^2}{4 \beta_\|^2}}  e^{\frac{-(p'_z-\bar{p}_z)^2}{4 \beta_\|^2}} e^{\frac{-(q_z-\bar{q}_z)^2}{4 \beta_\|^2}}  e^{\frac{-(p_z-\bar{p}_z)^2}{4 \beta_\|^2}}\nonumber\\
& \times& \sum_{s,r=1}^{2} \hat{a}^\dagger_{s}(p') \hat{a}_{s}(p) \hat{a}^\dagger_{r}(q') \hat{a}_{r}(q) \times~ [(p.q) (q'.p')]~.
\eea
Next, by integrating over $k$ and assuming that the center of the arm's tube is localized at $\bold{\bar{x}_\perp} = (W, 0)$ and $\bold{\bar{x}_\perp'} = (0, 0)$, according to Fig. \ref{ring}b, it becomes
\bea
\hat{H}_I(t)&=&\frac{ -\kappa^2 }{(16 \pi )^2} \int \frac{d^3\bold{p}}{(2\pi)^{3/2}}\int \frac{d^3\bold{p'}}{(2\pi)^{3/2}}
 \int \frac{d^3\bold{q}}{(2\pi)^{3/2}}\int \frac{d^3\bold{q'}}{(2\pi)^{3/2}} \int d{\bold{x}_\perp^2} \int d{\bold{x'}_\perp^2} \int_{-\infty} ^{\infty} dz \int_{-\infty} ^{\infty} dz' \nonumber\\
&\times & \frac{e^{i(p'^{0}-p^0+q^0-q'^0)t}}{\sqrt{p^0 q^0 p'^0 q'^0}} ~e^{-i(p'_z -p_z) (z-\bar{z})}~ e^{-i(q'_z-q_z) (z'-\bar{z}')} ~\times [(p.q) (q'.p')]\nonumber\\
&\times& \frac{1}{\sqrt{(\bold{x}_\perp - \bold{x'}_\perp)^2+ (z-z')^2}}~ e^{-i(\bold{p'-p})_\perp \cdot \bold{x}_\perp}~ e^{-i(\bold{q'-q})_\perp \cdot \bold{x'}_\perp} ~ e^{+i(p'_x-p_x) W} \sum_{s,r=1}^{2}\hat{a}^\dagger_{s}(p') \hat{a}_{s}(p) \hat{a}^\dagger_{r}(q') \hat{a}_{r}(q)\nonumber\\
&\times&\left(\frac{1}{2\pi \beta_\perp^2}\right)^2 \left(\frac{1}{2\pi \beta_{\|}^2}\right)~e^{\frac{-(\bold{q'}-\bold{\bar{q}})_\perp^2}{4 \beta_\perp^2}}  e^{\frac{-(\bold{p'}-\bold{\bar{p}})_\perp^2}{4 \beta_\perp^2}} e^{\frac{-(\bold{q}-\bold{\bar{q}})_\perp^2}{4 \beta_\perp^2}} e^{\frac{-(\bold{p}-\bold{\bar{p}})_\perp^2}{4 \beta_\perp^2}}~e^{\frac{-(q_z'-\bar{q}_z)^2}{4 \beta_\|^2}}  e^{\frac{-(p'_z-\bar{p}_z)^2}{4 \beta_\|^2}} e^{\frac{-(q_z-\bar{q}_z)^2}{4 \beta_\|^2}}  e^{\frac{-(p_z-\bar{p}_z)^2}{4 \beta_\|^2}}~,\nonumber\\
\eea
note that in the $z$ direction, the points ${\bar{z}}$ and ${\bar{z}'}$  vary within the range of $0$ to $L$, where $L$ refers to the lenght of the ring.
According to the $x$ and $y$ components of the tube's center, it reads
\bea
\hat{H}_I(t)&=&\frac{ -\kappa^2 }{(16 \pi )^2} \int \frac{d^3\bold{p}}{(2\pi)^{3/2}}\int \frac{d^3\bold{p'}}{(2\pi)^{3/2}}
 \int \frac{d^3\bold{q}}{(2\pi)^{3/2}}\int \frac{d^3\bold{q'}}{(2\pi)^{3/2}}  \int d{\bold{x}}_\perp^2 \int d \bold{x'}_\perp^2 \int_{-\infty} ^{\infty} dz \int_{-\infty} ^{\infty} dz' \nonumber\\
&\times &\left(\frac{1}{2\pi \beta_\perp^2}\right)^2 \left(\frac{1}{2\pi \beta_{\|}^2}\right) \frac{1}{\sqrt{p^0 q^0 p'^0 q'^0}}~e^{i(p'^{0}-p^0+q^0-q'^0)t} ~  ~e^{-i(p'_z -p_z) (z-\bar{z})}~ e^{-i(q'_z-q_z) (z'-\bar{z}')} ~\nonumber\\
&\times& e^{\frac{-(\bold{q'}-\bold{\bar{q}})_\perp^2}{4 \beta_\perp^2}}  e^{\frac{-(\bold{p'}-\bold{\bar{p}})_\perp^2}{4 \beta_\perp^2}} e^{\frac{-(\bold{q}-\bold{\bar{q}})_\perp^2}{4 \beta_\perp^2}} e^{\frac{-(\bold{p}-\bold{\bar{p}})_\perp^2}{4 \beta_\perp^2}}~e^{\frac{-(q_z'-\bar{q}_z)^2}{4 \beta_\|^2}}  e^{\frac{-(p'_z-\bar{p}_z)^2}{4 \beta_\|^2}} e^{\frac{-(q_z-\bar{q}_z)^2}{4 \beta_\|^2}}  e^{\frac{-(p_z-\bar{p}_z)^2}{4 \beta_\|^2}}\nonumber\\
& \times&\frac{e^{-i(p'_x-p_x) (x-W)}~e^{-i(p'_y-p_y) y}~ e^{-i(q'_x-q_x)  x'}~ e^{-i(q'_y-q_y)  y'}}{\sqrt{(\bold{x}_\perp - \bold{x'}_\perp)^2+ (z-z')^2}} \sum_{s,r=1}^{2}\hat{a}^\dagger_{s}(p') \hat{a}_{s}(p) \hat{a}^\dagger_{r}(q') \hat{a}_{r}(q) \times~ [(p.q) (q'.p')]~,
\eea
where we have used the signature $(+,-,-,-)$ of metric convention. By integrating over $\bold{p}_\perp'$ and $\bold{q}_\perp'$, this gives
\bea
\hat{H}_I(t)&=&\frac{ -\kappa^2 }{(16 \pi )^2} \int \frac{d^3\bold{p}}{(2\pi)^{3/2}}\int \frac{dp_z'}{(2\pi)^{3/2}}
 \int \frac{d^3\bold{q}}{(2\pi)^{3/2}}\int \frac{dq_z'}{(2\pi)^{3/2}}  \int d{\bold{x}_\perp^2} \int d{\bold{x'}_\perp ^2}  \int_{-\infty} ^{\infty} dz \int_{-\infty} ^{\infty} dz'  \nonumber\\
&\times & \left(\frac{1}{2\pi \beta_\perp^2}\right)^2 \left(\frac{1}{2\pi \beta_{\|}^2}\right)  \left(4\pi \beta_\perp^4\right)^2\frac{1}{\sqrt{p^0 q^0 p'^0 q'^0}}~ e^{i(p'^{0}-p^0+q^0-q'^0)t} ~  ~e^{-i(p'_z -p_z) (z-\bar{z})}~ e^{-i(q'_z-q_z) (z'-\bar{z}')} ~\nonumber\\
&\times& e^{\frac{-(\bold{q}-\bold{\bar{q}})_\perp^2}{4 \beta_\perp^2}} e^{\frac{-(\bold{p}-\bold{\bar{p}})_\perp^2}{4 \beta_\perp^2}}~e^{\frac{-(q_z'-\bar{q}_z)^2}{4 \beta_\|^2}}  e^{\frac{-(p'_z-\bar{p}_z)^2}{4 \beta_\|^2}} e^{\frac{-(q_z-\bar{q}_z)^2}{4 \beta_\|^2}}  e^{\frac{-(p_z-\bar{p}_z)^2}{4 \beta_\|^2}}~e^{-\beta_\perp^2 [(x-W)^2+y^2]}~e^{- \beta_\perp^2 [x'^2+y'^2]}\nonumber\\
&\times&\frac{e^{-i(\bar{p}_x-p_x) (x-W)}~e^{-i(\bar{p}_y-p_y) y}~ e^{-i(\bar{q}_x-q_x)  x'}~ e^{-i(\bar{q}_y-q_y)  y'}}{\sqrt{(\bold{x}_\perp - \bold{x'}_\perp)^2+ (z-z')^2}} \sum_{s,r=1}^{2}\hat{a}^\dagger_{s}(p') \hat{a}_{s}(p) \hat{a}^\dagger_{r}(q') \hat{a}_{r}(q)\nonumber\\
& \times& (p^0q^0-p_x q_x-p_y q_y- p_zq_z)\left\{4W x'-4(x x'+ yy')+\frac{2 i}{\beta_\perp^2} \left[-\bar{q}_x(x-W)-\bar{p}_x x'-\bar{q}_y y -\bar{p}_y y'\right]\right. \nonumber\\
&+&\left.\frac{1}{\beta_\perp^4}(p'^0 q'^0-\bar{p}_x \bar{q}_x-\bar{p}_y \bar{q}_y-p'_z q'_z)\right\}.
\eea
Following integrating over $\bold{p}_\perp$ and $\bold{q}_\perp$, we have
\bea
\hat{H}_I(t)&=&\frac{ -\kappa^2 }{(16 \pi )^2} \int \frac{dp_z}{(2\pi)^{3/2}}\int \frac{dp_z'}{(2\pi)^{3/2}}
 \int \frac{dq_z}{(2\pi)^{3/2}}\int \frac{dq_z'}{(2\pi)^{3/2}} \int_{-\infty} ^{\infty} dz \int_{-\infty} ^{\infty} dz' \sum_{s,r=1}^{2}\hat{a}^\dagger_{s}(p') \hat{a}_{s}(p) \hat{a}^\dagger_{r}(q') \hat{a}_{r}(q) \nonumber\\
&\times &\left(\frac{1}{2\pi \beta_\perp^2}\right)^2 \left(\frac{1}{2\pi \beta_{\|}^2}\right) \left(4\pi \beta_\perp^4\right)^4\frac{1}{\sqrt{p^0 q^0 p'^0 q'^0}}~  e^{i(p'^{0}-p^0+q^0-q'^0)t}~e^{-i(p'_z -p_z) (z-\bar{z})}~ e^{-i(q'_z-q_z) (z'-\bar{z}')} ~\nonumber\\
&\times&  \int d{\bold{x}_\perp^2} \int d{\bold{x'}_\perp ^2}~ \frac{e^{-2\beta_\perp^2 [(x-W)^2+y^2]}~e^{-2 \beta_\perp^2 [x'^2+y'^2]}}{\sqrt{(\bold{x}_\perp - \bold{x'}_\perp)^2+ (z-z')^2}} ~ e^{\frac{-(q_z'-\bar{q}_z)^2}{4\beta_\|^2}}  e^{\frac{-(p'_z-\bar{p}_z)^2}{4 \beta_\|^2}} e^{\frac{-(q_z-\bar{q}_z)^2}{4 \beta_\|^2}}  e^{\frac{-(p_z-\bar{p}_z)^2}{4 \beta_\|^2}}\nonumber\\
& \times&\left\{4W x'-4(x x'+ yy')-\frac{2 i}{ \beta_\perp^2} \left[\bar{q}_x(x-W)+\bar{p}_x x'+\bar{q}_y y +\bar{p}_y y'\right]+\frac{1}{\beta_\perp^4}(p'^0 q'^0-\bar{p}_x \bar{q}_x-\bar{p}_y \bar{q}_y-p'_z q'_z)\right\}\nonumber\\
& \times&\left\{4W x'-4(x x'+ yy')+\frac{2 i}{ \beta_\perp^2}  \left[\bar{q}_x(x-W)+\bar{p}_x x'+\bar{q}_y y +\bar{p}_y y'\right]+\frac{1}{\beta_\perp^4}(p^0 q^0-\bar{p}_x \bar{q}_x-\bar{p}_y \bar{q}_y-p_z q_z)\right\}.
\eea
According to Fig. \ref{ring}b, we have $\bar{p}_x=-\bar{q}_x$ and $\bar{p}_y=\bar{q}_y$.  Additionally, since the momenta in the $z$ direction in the two parallel arms are opposite, the Hamiltonian becomes
\bea
\hat{H}_I(t)&=&\frac{ -\kappa^2 }{(16 \pi )^2} \int \frac{dp_z}{(2\pi)^{3/2}}\int \frac{dp_z'}{(2\pi)^{3/2}}
 \int \frac{dq_z}{(2\pi)^{3/2}}\int \frac{dq_z'}{(2\pi)^{3/2}} \int_{-\infty} ^{\infty} dz \int_{-\infty} ^{\infty} dz'  \sum_{s,r=1}^{2} \hat{a}^\dagger_{s}(p') \hat{a}_{s}(p) \hat{a}^\dagger_{r}(q') \hat{a}_{r}(q)\nonumber\\
&\times &\left(\frac{1}{2\pi \beta_\perp^2}\right)^2 \left(\frac{1}{2\pi \beta_{\|}^2}\right) \left(4\pi \beta_\perp^4\right)^4 ~\frac{1}{\sqrt{p^0 q^0 p'^0 q'^0}}~  e^{i(p'^{0}-p^0+q^0-q'^0)t} ~  ~e^{-i(p'_z -p_z) (z-\bar{z})}~ e^{-i(q'_z-q_z) (z'-\bar{z}')} ~\nonumber\\
&\times&  \int d{\bold{x}_\perp^2} \int d{\bold{x'}_\perp^2}  ~ \frac{e^{-2\beta_\perp^2 [(x-W)^2+y^2]}~e^{-2 \beta_\perp^2 [x'^2+y'^2]}}{\sqrt{(\bold{x}_\perp - \bold{x'}_\perp)^2+ (z-z')^2}}~ e^{\frac{-(q_z'-\bar{q}_z)^2}{4\beta_\|^2}}  e^{\frac{-(p'_z-\bar{p}_z)^2}{4 \beta_\|^2}} e^{\frac{-(q_z-\bar{q}_z)^2}{4 \beta_\|^2}}  e^{\frac{-(p_z-\bar{p}_z)^2}{4 \beta_\|^2}} \nonumber\\
& \times& \left\{4W x'-4(x x'+ yy')-\frac{2 i} {\beta_\perp^2} \left[\bar{p}_x(x-W+x')+\bar{p}_y (y' - y)\right]+\frac{1}{\beta_\perp^4}(p'^0 q'^0+{\bar{p}_x}^2 -{\bar{p}_y}^2+|p'_z| |q'_z|)\right\}\nonumber\\
& \times& \left\{4W x'-4(x x'+ yy')+\frac{2 i} {\beta_\perp^2} \left[\bar{p}_x(x-W+x')+\bar{p}_y (y' - y)\right]+\frac{1}{\beta_\perp^4}(p^0 q^0+{\bar{p}_x}^2 -{\bar{p}_y}^2+|p_z| |q_z|)\right\},
\eea
therefore, assuming the symmetry in the $x$ and $y$ components of photon momenta, $|\bar{p}_x|=|\bar{p}_y|$, yields
\bea
\hat{H}_I(t)&=&\frac{ -\kappa^2 }{(16 \pi )^2} \int \frac{dp_z}{(2\pi)^{3/2}}\int \frac{dp_z'}{(2\pi)^{3/2}}
 \int \frac{dq_z}{(2\pi)^{3/2}}\int \frac{dq_z'}{(2\pi)^{3/2}} \int_{-\infty} ^{\infty} dz \int_{-\infty} ^{\infty} dz'  \sum_{s,r=1}^{2}\hat{a}^\dagger_{s}(p') \hat{a}_{s}(p) \hat{a}^\dagger_{r}(q') \hat{a}_{r}(q)\nonumber\\
&\times &\left(\frac{1}{2\pi \beta_\perp^2}\right)^2 \left(\frac{1}{2\pi \beta_{\|}^2}\right) \left(4\pi \beta_\perp^4\right)^4\frac{1}{\sqrt{p^0 q^0 p'^0 q'^0}}~  e^{i(p'^{0}-p^0+q^0-q'^0)t} ~  ~e^{-i(p'_z -p_z) (z-\bar{z})}~ e^{-i(q'_z-q_z) (z'-\bar{z}')} ~\nonumber\\
&\times&\int d{\bold{x}_\perp^2} \int d{\bold{x'}_\perp ^2}~ \frac{e^{-2\beta_\perp^2 [(x-W)^2+y^2]}~e^{-2 \beta_\perp^2 [x'^2+y'^2]} }{\sqrt{(\bold{x}_\perp - \bold{x'}_\perp)^2+ (z-z')^2}}~e^{\frac{-(q_z'-\bar{q}_z)^2}{4\beta_\|^2}}  e^{\frac{-(p'_z-\bar{p}_z)^2}{4 \beta_\|^2}} e^{\frac{-(q_z-\bar{q}_z)^2}{4 \beta_\|^2}}  e^{\frac{-(p_z-\bar{p}_z)^2}{4 \beta_\|^2}} \nonumber\\
& \times& \left\{4W x'-4(x x'+ yy')-\frac{2 i}{ \beta_\perp^2} \bar{p}_x(x-W+x'-y + y') +\frac{1}{\beta_\perp^4}(p'^0 q'^0+|p'_z| |q'_z|)\right\}\nonumber\\
& \times& \left\{4W x'-4(x x'+ yy')+\frac{2 i}{ \beta_\perp^2} \bar{p}_x(x-W+x'-y + y')+\frac{1}{\beta_\perp^4}(p^0 q^0+|p_z| |q_z|)\right\}~.
\eea
After simplifying and substituting $\beta_\perp=1/\sigma$, we have
\bea
\hat{H}_I(t)&=&\frac{ -\kappa^2 }{(16 \pi )^2} \int \frac{dp_z}{(2\pi)^{3/2}}\int \frac{dp_z'}{(2\pi)^{3/2}}
 \int \frac{dq_z}{(2\pi)^{3/2}}\int \frac{dq_z'}{(2\pi)^{3/2}} \int_{-\infty} ^{\infty} dz \int_{-\infty} ^{\infty} dz' \sum_{s,r=1}^{2}\hat{a}^\dagger_{s}(p') \hat{a}_{s}(p) \hat{a}^\dagger_{r}(q') \hat{a}_{r}(q)\nonumber\\
&\times &  \left(\frac{\sigma^2}{2\pi}\right)^2 \left(\frac{1}{2\pi \beta_{\|}^2}\right)  \left(\frac{4\pi}{ \sigma^4}\right)^4 \frac{1}{\sqrt{p^0 q^0 p'^0 q'^0}}~ e^{i(p'^{0}-p^0+q^0-q'^0)t} ~  ~e^{-i(p'_z -p_z) (z-\bar{z})}~ e^{-i(q'_z-q_z) (z'-\bar{z}')} \nonumber\\
& \times&\int d{\bold{x}_\perp^2} \int d{\bold{x'}_\perp^2}~\frac{e^{-\frac{2}{\sigma^2} [(x-W)^2+y^2]}~e^{-\frac{2}{ \sigma^2} [x'^2+y'^2]}}{\sqrt{(\bold{x}_\perp - \bold{x'}_\perp)^2+ (z-z')^2}} ~e^{\frac{-(q_z'-\bar{q}_z)^2}{4 \beta_\|^2}}  e^{\frac{-(p'_z-\bar{p}_z)^2}{4 \beta_\|^2}} e^{\frac{-(q_z-\bar{q}_z)^2}{4 \beta_\|^2}}  e^{\frac{-(p_z-\bar{p}_z)^2}{4 \beta_\|^2}}\nonumber\\
& \times&\left\{[4W x'-4(x x'+ yy')+{\sigma^4}(p'^0 q'^0+|p'_z| |q'_z|)]^2+{4  {\bar{p}_x}^2 \sigma^4} (x-W+x'-y+y')^2\right\}  . \nonumber\\ \label{Gaussian-gravity}
\eea
Here, the exact calculation of this Gaussian integral over $\bold{x}_\perp$ and $\bold{x}'_\perp$ is not possible. One of the ways we have approached this Gaussian integral is through the discrete approximation of the Gaussian distribution using the error function. Thus, Eq. \eqref{Gaussian-gravity}, along with the approximation of the discrete Gaussian distribution (Fig. \ref{discrete}), yields

\bea
\hat{H}_I(t)&\simeq&\frac{ -\kappa^2 }{(16 \pi )^2} \int \frac{dp_z}{(2\pi)^{3/2}}\int \frac{dp_z'}{(2\pi)^{3/2}}
 \int \frac{dq_z}{(2\pi)^{3/2}}\int \frac{dq_z'}{(2\pi)^{3/2}} \int_{-\infty} ^\infty dz \int_{-\infty} ^\infty dz'~ \left(\frac{\sigma^2}{2\pi}\right)^2  \left(\frac{4\pi}{\sigma^4}\right)^4\left(\frac{\pi \sigma^2}{2}\right)^2 \nonumber\\
& \times& \int d{\bold{x}_\perp^2} \int d{\bold{x'}_\perp^2}~\frac{1}{\sqrt{(\bold{x}_\perp - \bold{x'}_\perp)^2+ (z-z')^2}} ~\frac{e^{i(p'^{0}-p^0+q^0-q'^0)t} }{\sqrt{p^0 q^0 p'^0 q'^0}}~\sum_{s,r=1}^{2}\hat{a}^\dagger_{s}(p') \hat{a}_{s}(p) \hat{a}^\dagger_{r}(q') \hat{a}_{r}(q) \nonumber\\
&\times&\left(\frac{1}{2\pi \beta_{\|}^2}\right) e^{\frac{-(q_z'-\bar{q}_z)^2}{4 \beta_\|^2}}  e^{\frac{-(p'_z-\bar{p}_z)^2}{4 \beta_\|^2}} e^{\frac{-(q_z-\bar{q}_z)^2}{4 \beta_\|^2}}  e^{\frac{-(p_z-\bar{p}_z)^2}{4 \beta_\|^2}}~e^{-i(p'_z -p_z) (z-\bar{z})}~ e^{-i(q'_z-q_z) (z'-\bar{z}')} \nonumber\\
&\times&\left\{P_{0.1\sigma} \delta(x-W)+P_{+\sigma} \delta(x-W-\sigma)+P_{-\sigma} \delta(x-W+\sigma)+...\right\}\nonumber\\
&\times&\left\{ P_{0.1\sigma} \delta(y) +P_{+\sigma} \delta(y-\sigma)+P_{-\sigma} \delta(y+\sigma)+...\right\}\nonumber\\
&\times&\left\{ P_{0.1\sigma} \delta(x') +P_{+\sigma} \delta(x'-\sigma)+P_{-\sigma}  \delta(x'+\sigma)+...\right\}\nonumber\\
&\times&\left\{ P_{0.1\sigma} \delta(y') +P_{+\sigma} \delta(y'-\sigma)+P_{-\sigma} \delta(y'+\sigma)+...\right\}\nonumber\\
& \times&\left\{[4W x'-4(x x'+ yy')+\sigma^4(p'^0 q'^0+|p'_z| |q'_z|)]^2+4 \sigma^4 {\bar{p}_x}^2 (x-W+x'-y+y')^2\right\}, \label{approx-hamilton}
\eea
where we use
\bea
P_{a-b}=P(a \le x \le b)=P\left(\frac{a-\mu}{\sigma} \le Z \le \frac{b-\mu}{\sigma}\right)=\Phi \left(\frac{b-\mu}{\sigma}\right)-\Phi \left(\frac{a-\mu}{\sigma}\right)~,
\eea
and for a normal distribution, we have the cumulative distribution function
\bea
\Phi(x)=P(Z \le x)=\frac{1}{\sqrt{2 \pi}} \int_{- \infty}^x \text{exp}\left(-\frac{u^2}{2}\right) du~,
\eea
which is related to the error function, erf($x$), as
\bea
\text{erf}(x)=2\Phi(\sqrt{2}x)-1~.
\eea
Therefore, \eqref{approx-hamilton} becomes
\bea
\hat{H}_I(t)&\simeq&\frac{ -\kappa^2 }{(16 \pi )^2} \int \frac{dp_z}{(2\pi)^{3/2}}\int \frac{dp_z'}{(2\pi)^{3/2}}
 \int \frac{dq_z}{(2\pi)^{3/2}}\int \frac{dq_z'}{(2\pi)^{3/2}} \int_{-\infty} ^\infty dz \int_{-\infty} ^\infty dz'~ \left(\frac{\sigma^2}{2\pi}\right)^2  \left(\frac{4\pi}{\sigma^4}\right)^4\left(\frac{\pi \sigma^2}{2}\right)^2 \nonumber\\
& \times& \int d{\bold{x}_\perp^2} \int d{\bold{x'}_\perp^2}~\frac{1}{\sqrt{(\bold{x}_\perp - \bold{x'}_\perp)^2+ (z-z')^2}} ~\frac{e^{i(p'^{0}-p^0+q^0-q'^0)t}}{\sqrt{p^0 q^0 p'^0 q'^0}}~\sum_{s,r=1}^2 \hat{a}^\dagger_{s}(p') \hat{a}_{s}(p) \hat{a}^\dagger_{r}(q') \hat{a}_{r}(q) \nonumber\\
&\times&\left(\frac{1}{2\pi \beta_{\|}^2}\right)  e^{\frac{-(q_z'-\bar{q}_z)^2}{4 \beta_\|^2}}  e^{\frac{-(p'_z-\bar{p}_z)^2}{4 \beta_\|^2}} e^{\frac{-(q_z-\bar{q}_z)^2}{4 \beta_\|^2}}  e^{\frac{-(p_z-\bar{p}_z)^2}{4 \beta_\|^2}}~e^{-i(p'_z -p_z) (z-\bar{z})}~ e^{-i(q'_z-q_z) (z'-\bar{z}')} \nonumber\\
&\times&\left\{(0.1) \delta(x-W)+(0.449) \delta(x-W-\sigma)+(0.449) \delta(x-W+\sigma)+...\right\}\nonumber\\
&\times&\left\{ (0.1) \delta(y) +(0.449) \delta(y-\sigma)+(0.449) \delta(y+\sigma)+...\right\}\nonumber\\
&\times&\left\{ (0.1) \delta(x') +(0.449) \delta(x'-\sigma)+(0.449) \delta(x'+\sigma)+...\right\}\nonumber\\
&\times&\left\{ (0.1) \delta(y') +(0.449) \delta(y'-\sigma)+(0.449) \delta(y'+\sigma)+...\right\}\nonumber\\
& \times&\left\{[4W x'-4(x x'+ yy')+\sigma^4(p'^0 q'^0+|p'_z| |q'_z|)]^2+4 \sigma^4 {\bar{p}_x}^2 (x-W+x'-y+y')^2\right\}.
\eea
By integrating over the perpendicular directions using these Dirac delta functions, we have
\bea
\hat{H}_I(t)&\simeq&\frac{ -\kappa^2 }{(16 \pi )^2} \int \frac{dp_z}{(2\pi)^{3/2}}\int \frac{dp_z'}{(2\pi)^{3/2}}
 \int \frac{dq_z}{(2\pi)^{3/2}}\int \frac{dq_z'}{(2\pi)^{3/2}}  \frac{e^{i(p'^{0}-p^0+q^0-q'^0)t}}{\sqrt{p^0 q^0 p'^0 q'^0}} \sum_{s,r=1}^2  \hat{a}^\dagger_{s}(p') \hat{a}_{s}(p) \hat{a}^\dagger_{r}(q') \hat{a}_{r}(q)\nonumber\nonumber\\
&\times & \left(\frac{1}{2\pi \beta_{\|}^2}\right)~ e^{\frac{-(q_z'-\bar{q}_z)^2}{4 \beta_\|^2}}  e^{\frac{-(p'_z-\bar{p}_z)^2}{4 \beta_\|^2}} e^{\frac{-(q_z-\bar{q}_z)^2}{4 \beta_\|^2}}  e^{\frac{-(p_z-\bar{p}_z)^2}{4 \beta_\|^2}}~e^{-i(p'_z -p_z) (z-\bar{z})}~ e^{-i(q'_z-q_z) (z'-\bar{z}')} \nonumber\\
&\times&\int_{-\infty} ^\infty dz \int_{-\infty} ^\infty dz'~\left(\frac{16\pi^4}{\sigma^4}\right)~\left\{\sigma^{4}(p'^0 q'^0+|p'_z| |q'_z|)^2 \left[\frac{ 0.172}{\sqrt{W^2+ (z-z')^2}}+\frac{ 0.037}{\sqrt{(W-\sigma)^2+ (z-z')^2}}\right. \right.\nonumber\\
&+&\left. \left.\frac{ 0.037}{\sqrt{(W+\sigma)^2+ (z-z')^2}}+\frac{ 0.075}{\sqrt{W^2+\sigma^2+ (z-z')^2}} +\frac{ 0.084}{\sqrt{(W-2\sigma)^2+ (z-z')^2}}+\frac{ 0.084}{\sqrt{(W+2\sigma)^2+ (z-z')^2}} \right. \right.\nonumber\\
&+& \left. \left.\frac{ 0.168}{\sqrt{W^2+4\sigma^2+ (z-z')^2}}+\frac{ 0.016}{\sqrt{W^2+2\sigma^2+2 W \sigma+ (z-z')^2}}+\frac{ 0.016}{\sqrt{W^2+2\sigma^2-2 W \sigma+ (z-z')^2}} \right.\right.\nonumber\\
&+& \left. \left.\frac{ 0.036}{\sqrt{W^2+5\sigma^2+4 W \sigma+ (z-z')^2}}+\frac{ 0.036}{\sqrt{W^2+5\sigma^2-4 W \sigma+ (z-z')^2}}+\frac{ 0.036}{\sqrt{W^2+5\sigma^2-2 W \sigma+ (z-z')^2}} \right.\right.\nonumber\\
&+& \left. \left.\frac{ 0.036}{\sqrt{W^2+5\sigma^2+2 W \sigma+ (z-z')^2}}+\frac{ 0.082}{\sqrt{W^2+8\sigma^2+4 W \sigma+ (z-z')^2}}+\frac{ 0.082}{\sqrt{W^2+8\sigma^2-4 W \sigma+ (z-z')^2}} \right]\right.\nonumber\\
&+&\left.4 \sigma^{2}(p'^0 q'^0+|p'_z| |q'_z|)\left[-\frac{ 0.672}{\sqrt{W^2+ (z-z')^2}}-\frac{ 0.073}{\sqrt{(W-\sigma)^2+ (z-z')^2}}-\frac{ 0.073}{\sqrt{(W+\sigma)^2+ (z-z')^2}}\right. \right.\nonumber\\
&-&\left. \left.\frac{ 0.146}{\sqrt{W^2+\sigma^2+ (z-z')^2}} +\frac{ 0.004}{\sqrt{(W-2\sigma)^2+ (z-z')^2}}+\frac{ 0.004}{\sqrt{(W+2\sigma)^2+ (z-z')^2}}+\frac{ 0.008}{\sqrt{W^2+4\sigma^2+ (z-z')^2}} \right.\right.\nonumber\\
&+&\left. \left.\frac{ 0.073}{\sqrt{W^2+5\sigma^2+2W \sigma+ (z-z')^2}}+\frac{ 0.073}{\sqrt{W^2+5\sigma^2-2W \sigma+ (z-z')^2}} +\frac{ 0.073}{\sqrt{W^2+5\sigma^2+4W \sigma+ (z-z')^2}} \right.\right.\nonumber\\
&+&\left. \left.\frac{ 0.073}{\sqrt{W^2+5\sigma^2-4W \sigma+ (z-z')^2}}+\frac{ 0.328}{\sqrt{W^2+8\sigma^2-4W \sigma+ (z-z')^2}}+\frac{ 0.328}{\sqrt{W^2+8\sigma^2+4W \sigma+ (z-z')^2}} \right]\right.\nonumber\\
&+&\left. 4 \sigma^{2}\bar{p}_x^2\left[\frac{ 0.672}{\sqrt{W^2+ (z-z')^2}}+\frac{ 0.037}{\sqrt{(W-\sigma)^2+ (z-z')^2}}+\frac{ 0.037}{\sqrt{(W+\sigma)^2+ (z-z')^2}}+\frac{ 0.366}{\sqrt{W^2+\sigma^2+ (z-z')^2}} \right. \right.\nonumber\\
&+&\left. \left.\frac{ 0.032}{\sqrt{(W-\sigma)^2+\sigma^2+ (z-z')^2}}+\frac{ 0.032}{\sqrt{(W+\sigma)^2+\sigma^2+ (z-z')^2}} +\frac{ 0.366}{\sqrt{W^2+4\sigma^2+ (z-z')^2}} +\cdot \cdot \cdot \right]\right.\nonumber\\
&+& \left.  16\left[\frac{ 0.664}{\sqrt{W^2+ (z-z')^2}}+\frac{ 0.036}{\sqrt{(W-\sigma)^2+ (z-z')^2}}+\frac{ 0.036}{\sqrt{(W+\sigma)^2+ (z-z')^2}}+\frac{ 0.073}{\sqrt{W^2+\sigma^2+ (z-z')^2}}\right. \right.\nonumber\\
&+&\left. \left.\frac{ 0.002}{\sqrt{(W-2\sigma)^2+ (z-z')^2}}+\frac{ 0.002}{\sqrt{(W+2\sigma)^2+ (z-z')^2}} +\cdot \cdot \cdot \right]\right\}. \nonumber\\ \label{GHamilton}
\eea
Finally, after integrating over the momenta in the $z$ direction, since the points ${\bar{z}}$ and ${\bar{z}'}$ extend from $0$ to $L$, the integrals concerning the momenta of the $z$ component result in the rectangular function \(\Pi_0^L(z)\) \cite{Cywiak:2010, Hsu:1970}. Assuming a single mode cavity with $|\bar{p}_z|=|\bar{q}_z|={p_0}$, we obtain
\bea
\hat{H}_I(t)&\simeq&\frac{ -\kappa^2 }{(16 \pi )^2} \frac{1}{{{p^0}^2}} \frac{1}{\pi^3 L^2} \sum_{s,r=1}^2  \hat{a}^\dagger_{s}({p_0})~ \hat{a}_{s}({p_0}) ~\hat{a}^\dagger_{r}(-{p_0}) ~\hat{a}_{r}(-{p_0})
 \nonumber\\
&\times&\int_{0} ^L dz \int_{0} ^L dz'~\left(\frac{2\pi}{\sigma}\right)^4~\left\{4 \sigma^{4} {{p_0}}^4 \left[\frac{ 0.172}{\sqrt{W^2+ (z-z')^2}}+\frac{ 0.037}{\sqrt{(W-\sigma)^2+ (z-z')^2}}\right. \right.\nonumber\\
&+&\left. \left.\frac{ 0.037}{\sqrt{(W+\sigma)^2+ (z-z')^2}}+\frac{ 0.075}{\sqrt{W^2+\sigma^2+ (z-z')^2}} +\frac{ 0.084}{\sqrt{(W-2\sigma)^2+ (z-z')^2}}+\frac{ 0.084}{\sqrt{(W+2\sigma)^2+ (z-z')^2}} \right. \right.\nonumber\\
&+& \left. \left.\frac{ 0.168}{\sqrt{W^2+4\sigma^2+ (z-z')^2}}+\frac{ 0.016}{\sqrt{W^2+2\sigma^2+2 W \sigma+ (z-z')^2}}+\frac{ 0.016}{\sqrt{W^2+2\sigma^2-2 W \sigma+ (z-z')^2}} \right.\right.\nonumber\\
&+& \left. \left.\frac{ 0.036}{\sqrt{W^2+5\sigma^2+4 W \sigma+ (z-z')^2}}+\frac{ 0.036}{\sqrt{W^2+5\sigma^2-4 W \sigma+ (z-z')^2}}+\frac{ 0.036}{\sqrt{W^2+5\sigma^2-2 W \sigma+ (z-z')^2}} \right.\right.\nonumber\\
&+& \left. \left.\frac{ 0.036}{\sqrt{W^2+5\sigma^2+2 W \sigma+ (z-z')^2}}+\frac{ 0.082}{\sqrt{W^2+8\sigma^2+4 W \sigma+ (z-z')^2}}+\frac{ 0.082}{\sqrt{W^2+8\sigma^2-4 W \sigma+ (z-z')^2}} \right]\right.\nonumber\\
&+&\left. 8\sigma^{2}{p_0 }^2\left[-\frac{ 0.672}{\sqrt{W^2+ (z-z')^2}}-\frac{ 0.073}{\sqrt{(W-\sigma)^2+ (z-z')^2}}-\frac{ 0.073}{\sqrt{(W+\sigma)^2+ (z-z')^2}}\right. \right.\nonumber\\
&-&\left. \left.\frac{ 0.146}{\sqrt{W^2+\sigma^2+ (z-z')^2}} +\frac{ 0.004}{\sqrt{(W-2\sigma)^2+ (z-z')^2}}+\frac{ 0.004}{\sqrt{(W+2\sigma)^2+ (z-z')^2}}+\frac{ 0.008}{\sqrt{W^2+4\sigma^2+ (z-z')^2}} \right.\right.\nonumber\\
&+&\left. \left.\frac{ 0.073}{\sqrt{W^2+5\sigma^2+2W \sigma+ (z-z')^2}}+\frac{ 0.073}{\sqrt{W^2+5\sigma^2-2W \sigma+ (z-z')^2}} +\frac{ 0.073}{\sqrt{W^2+5\sigma^2+4W \sigma+ (z-z')^2}} \right.\right.\nonumber\\
&+&\left. \left.\frac{ 0.073}{\sqrt{W^2+5\sigma^2-4W \sigma+ (z-z')^2}}+\frac{ 0.328}{\sqrt{W^2+8\sigma^2-4W \sigma+ (z-z')^2}}+\frac{ 0.328}{\sqrt{W^2+8\sigma^2+4W \sigma+ (z-z')^2}} \right]\right.\nonumber\\
&+&\left.4\sigma^{2}\bar{p}_x^2\left[\frac{ 0.672}{\sqrt{W^2+ (z-z')^2}}+\frac{ 0.037}{\sqrt{(W-\sigma)^2+ (z-z')^2}}+\frac{ 0.037}{\sqrt{(W+\sigma)^2+ (z-z')^2}}+\frac{ 0.366}{\sqrt{W^2+\sigma^2+ (z-z')^2}} \right. \right.\nonumber\\
&+&\left. \left.\frac{ 0.032}{\sqrt{(W-\sigma)^2+\sigma^2+ (z-z')^2}}+\frac{ 0.032}{\sqrt{(W+\sigma)^2+\sigma^2+ (z-z')^2}} +\frac{ 0.366}{\sqrt{W^2+4\sigma^2+ (z-z')^2}} +\cdot \cdot \cdot \right]\right.\nonumber\\
&+& \left.  16\left[\frac{ 0.664}{\sqrt{W^2+ (z-z')^2}}+\frac{ 0.036}{\sqrt{(W-\sigma)^2+ (z-z')^2}}+\frac{ 0.036}{\sqrt{(W+\sigma)^2+ (z-z')^2}}+\frac{ 0.073}{\sqrt{W^2+\sigma^2+ (z-z')^2}}\right. \right.\nonumber\\
&+&\left. \left.\frac{ 0.002}{\sqrt{(W-2\sigma)^2+ (z-z')^2}}+\frac{ 0.002}{\sqrt{(W+2\sigma)^2+ (z-z')^2}} +\cdot \cdot \cdot \right]\right\}~. \label{GHamilton}
\eea
The more straightforward expression of this equation can be formulated as 
\bea
\hat{H}_I(t)&\simeq&\frac{ -4G }{{p^0}^2 L^2 \sigma^4} \left\{ \sigma^{4} {{p_0}}^4 \mathcal{A}_0(\sigma)+2\sigma^{2}{p_0 }^2 \mathcal{A}_1(\sigma)+\sigma^{2}\bar{p}_x^2\mathcal{A}_2(\sigma)+4\mathcal{A}_3(\sigma)\right\}  \sum_{s,r=1}^2  \hat{a}^\dagger_{s}({p_0})~ \hat{a}_{s}({p_0}) ~\hat{a}^\dagger_{r}(-{p_0}) ~\hat{a}_{r}(-{p_0})~,\nonumber\\ \label{GHamilton-bfinal}
\eea
where
\bea
\mathcal{A}_0(\sigma)= \int_0 ^L dz \int_0 ^L  dz' \left[\frac{ 0.172}{\sqrt{W^2+ (z-z')^2}}+\frac{ 0.037}{\sqrt{(W-\sigma)^2+ (z-z')^2}}+\frac{ 0.037}{\sqrt{(W+\sigma)^2+ (z-z')^2}}+\cdot \cdot \cdot \right]~,
\eea
\bea
\mathcal{A}_1(\sigma)= \int_0 ^L dz \int_0 ^L  dz' \left[-\frac{ 0.672}{\sqrt{W^2+ (z-z')^2}}-\frac{ 0.073}{\sqrt{(W-\sigma)^2+ (z-z')^2}}-\frac{ 0.073}{\sqrt{(W+\sigma)^2+ (z-z')^2}}+\cdot \cdot \cdot \right]~,
\eea
\bea
\mathcal{A}_2(\sigma)= \int_0 ^L dz \int_0 ^L  dz' \left[\frac{ 0.672}{\sqrt{W^2+ (z-z')^2}}+\frac{ 0.037}{\sqrt{(W-\sigma)^2+ (z-z')^2}}+\frac{ 0.037}{\sqrt{(W+\sigma)^2+ (z-z')^2}}+\cdot \cdot \cdot \right]~,
\eea
\bea
\mathcal{A}_3(\sigma)= \int_0 ^L dz \int_0 ^L  dz' \left[\frac{ 0.664}{\sqrt{W^2+ (z-z')^2}}+\frac{ 0.036}{\sqrt{(W-\sigma)^2+ (z-z')^2}}+\frac{ 0.036}{\sqrt{(W+\sigma)^2+ (z-z')^2}}+\cdot \cdot \cdot \right]~.
\eea
Note that in Eq. \eqref{GHamilton-bfinal}, the third term is negligible compared to the second, as $|\bar{p}_x|$ can be ignored relative to $|\bar{p}_z|=p_0$ under a geometric optics condition where $L \gg \sigma$. Thus, for more simplicity, Eq. \eqref{GHamilton-bfinal}  can be expressed as
\bea
\hat{H}_I(t)&\simeq&\frac{ -4G }{{p^0}^2 L^2 \sigma^4} \left\{ \sigma^{4} {{p_0}}^4 \mathcal{A}_0(\sigma)+2\sigma^{2}{p_0 }^2 \mathcal{A}_1(\sigma)+4\mathcal{A}_3(\sigma)\right\}  \sum_{s,r=1}^2  \hat{a}^\dagger_{s}({p_0})~ \hat{a}_{s}({p_0}) ~\hat{a}^\dagger_{r}(-{p_0}) ~\hat{a}_{r}(-{p_0})~.  \label{GHamilton-final}
\eea
Hereafter, the $z$ direction integral will be calculated numerically.
\section{Hamiltonian of photons self-interaction with axion mediator}
 \label{axionapp}
Starting with the Hamiltonian in Eq. \eqref{axi-Hamiltoni}, we have
\bea
\hat{H}_I (t)&=&\frac{ g^2_{a\gamma \gamma} }{4} \int d^4 x' \int d^3\bold{x} \int \frac{d^3\bold{p}}{(2\pi)^{3/2}}\frac{1}{\sqrt{2p^0}} \int \frac{d^3\bold{p'}}{(2\pi)^{3/2}} \frac{1}{\sqrt{2p'^0}} \int \frac{d^3\bold{q}}{(2\pi)^{3/2}}\frac{1}{\sqrt{2q^0}} \int \frac{d^3\bold{q'}}{(2\pi)^{3/2}}\frac{1}{\sqrt{2q'^0}}\int \frac{d^4k}{(2\pi)^4} \nonumber\\
&\times &\left(\frac{1}{2\pi \beta_{\|}^2}\right) \left(\frac{1}{2\pi \beta_\perp^2}\right)^2\sum_{s , s' =1}^{2} \sum_{r,r' =1}^{2} e^{i(p'-p).(x-\bar{x})} e^{-i(q-q').(x'-\bar{x}')}  \frac{i e^{-ik.(x-x')}}{k^2-m_a^2+ik^0\Gamma_\beta(k^0)} \nonumber \\
&\times& e^{\frac{-(q_z'-\bar{q}_z)^2}{4 \beta_\|^2}}  e^{\frac{-(p'_z-\bar{p}_z)^2}{4 \beta_\|^2}} e^{\frac{-(q_z-\bar{q}_z)^2}{4 \beta_\|^2}}  e^{\frac{-(p_z-\bar{p}_z)^2}{4 \beta_\|^2}}~ e^{\frac{-(\bold{q'}-\bold{\bar{q}})^2}{4 \beta_\perp^2 }}  e^{\frac{-(\bold{p'}-\bold{\bar{p}})^2}{4 \beta_\perp^2}} e^{\frac{-(\bold{q}-\bold{\bar{q}})^2}{4 \beta_\perp^2}} e^{\frac{-(\bold{p}-\bold{\bar{p}})^2}{4 \beta_\perp^2}}~\nonumber\\
&\times& \{[(p.q') (q.p')- (p.q) (p'.q')]  [(\epsilon_r(q).\epsilon_s(p)) (\epsilon_{r'}(q').\epsilon_{s'}(p'))- (\epsilon_r(q).\epsilon_{s'}(p')) (\epsilon_{r'}(q').\epsilon_s(p))]\} \label{Hamiltoni}\nonumber\\
&\times&  \hat{a}^\dagger_{s'}(p') \hat{a}_{s}(p) \hat{a}^\dagger_{r'}(q') \hat{a}_{r}(q)~.
\eea
Then, after integrating over $t'$, it becomes
\bea
\hat{H}_I (t)&=&\frac{ g^2_{a\gamma \gamma} }{16} \int d^3 x' \int d^3\bold{x} \int \frac{d^3\bold{p}}{(2\pi)^{3/2}} \int \frac{d^3\bold{p'}}{(2\pi)^{3/2}}\int \frac{d^3\bold{q}}{(2\pi)^{3/2}} \int \frac{d^3\bold{q'}}{(2\pi)^{3/2}} \int \frac{d^4{k}}{(2\pi)^4}~\frac{2\pi   \delta (q'^{0}-q^0+k^0)}{\sqrt{p^0 q^0 p'^0 q'^0}}\nonumber\\
&\times &\sum_{s , s' =1}^{2} \sum_{r,r' =1}^{2} e^{i(p'^{0}-p^0-k^0)t}~e^{i(p'-p).(x-\bar{x})}~ e^{-i(q-q').(x'-\bar{x}')}~ \frac{i e^{-i \bold{k}.(\bold{x-x'})}}{{k^0}^2-\bold{k}^2-m_a^2+ik^0\Gamma_\beta(k^0)} \nonumber\\
&\times& \left(\frac{1}{2\pi \beta_{\|}^2}\right)\left(\frac{1}{2\pi \beta_\perp^2}\right)^2~e^{\frac{-(q_z'-\bar{q}_z)^2}{4 \beta_\|^2}}  e^{\frac{-(p'_z-\bar{p}_z)^2}{4 \beta_\|^2}} e^{\frac{-(q_z-\bar{q}_z)^2}{4 \beta_\|^2}}  e^{\frac{-(p_z-\bar{p}_z)^2}{4 \beta_\|^2}} e^{\frac{-(\bold{q'}-\bold{\bar{q}})^2}{4 \beta_\perp^2 }}  e^{\frac{-(\bold{p'}-\bold{\bar{p}})^2}{4 \beta_\perp^2}} e^{\frac{-(\bold{q}-\bold{\bar{q}})^2}{4 \beta_\perp^2}} e^{\frac{-(\bold{p}-\bold{\bar{p}})^2}{4 \beta_\perp^2}}\nonumber\\
&\times& \{[(p.q') (q.p')- (p.q) (p'.q')]  [(\epsilon_r(q).\epsilon_s(p)) (\epsilon_{r'}(q').\epsilon_{s'}(p')) - (\epsilon_r(q).\epsilon_{s'}(p')) (\epsilon_{r'}(q').\epsilon_s(p))]\} \label{Hamiltoni}\nonumber\\
& \times& \hat{a}^\dagger_{s'}(p') \hat{a}_{s}(p) \hat{a}^\dagger_{r'}(q') \hat{a}_{r}(q)~.
\eea
Hence, integrating over perpendicular momenta, yields
\bea
\hat{H}_I (t)&=&\frac{ g^2_{a\gamma \gamma} }{16} \left(\frac{1}{2\pi \beta_\perp^2}\right)^2  \left(\frac{1}{2\pi \beta_{\|}^2}\right) \left(16\pi^2 \beta_\perp^6\right)^2~ \int \frac{dp_z}{(2\pi)^{3/2}} \int \frac{dp'_z}{(2\pi)^{3/2}}~ \int \frac{dq_z}{(2\pi)^{3/2}}\int \frac{dq'_z}{(2\pi)^{3/2}}\int \frac{d^3\bold{k}}{(2\pi)^{3}} \nonumber\\
&\times&  \frac{e^{i(p'^{0}-p^0+q^0-q'^0)t}}{\sqrt{p^0 q^0 p'^0 q'^0}} \int d^3\bold{x'} \int d^3\bold{x} ~e^{-2\beta_\perp^2[(x-W)^2+y^2]}~e^{-2\beta_\perp^2[x'^2+y'^2]} \frac{i e^{-i\bold{k}.(\bold{x-x'})}}{(q^0-q'^0)^2-\bold{k}^2-m_a^2+ik^0\Gamma_\beta(k^0)}\nonumber\\
&\times&e^{\frac{-(q_z'-\bar{q}_z)^2}{4 \beta_\|^2}}  e^{\frac{-(p'_z-\bar{p}_z)^2}{4 \beta_\|^2}} e^{\frac{-(q_z-\bar{q}_z)^2}{4 \beta_\|^2}}  e^{\frac{-(p_z-\bar{p}_z)^2}{4 \beta_\|^2}}~e^{i(p'_z-p_z).(z-\bar{z})} ~e^{-i(q_z-q'_z).(z'-\bar{z}')}~\nonumber\\
&\times&\left\{4 x' (x-W)(q'^0 p^0+|q'_z| |p_z|+q^0 p'^0+|q_z| |p'_z|+q^0 p^0+|q_z| |p_z|+p'^0 q'^0+|p'_z| |q'_z|-4 \bar{p_y}^2)+16\bar{p_x}\bar{p_y} x' y\right. \nonumber\\ 
&+&\left.4 y'\left[y(q'^0 p^0+|q'_z| |p_z|+q^0 p'^0+|q_z| |p'_z|+q^0 p^0+|q_z| |p_z|+p'^0 q'^0+|p'_z| |q'_z|+4 \bar{p_x}^2)+4\bar{p_x}\bar{p_y} (W-x)\right] \right. \nonumber\\
&+&\left. \frac{2 i}{ \beta_\perp^2}\left[-(q^0 p^0+|q_z| |p_z|-p'^0 q'^0+|p'_z| |q'_z|) \left[\bar{p}_x(W-x-x')+\bar{p}_y(y-y') \right]+(q'^0 p^0+|q'_z| |p_z|)\right. \right. \nonumber\\
&\times& \left. \left. \left[\bar{p_x}(W-x+x')+\bar{p_y}(y+y') \right]-(q^0 p'^0+|q_z| |p'_z|)\left[\bar{p}_x(W-x+x')+\bar{p}_y(y+y')\right]\right]  \right. \nonumber\\
 &+&  \left.\frac{1}{\beta_\perp^4} \left[-(p^0q^0+|p_z| |q_z|) (p'^0 q'^0+|p'_z| |q'_z|)-(p^0q^0+|p_z| |q_z|+p'^0 q'^0+|p'_z| |q'_z|-q^0 p'^0+|q_z| |p'_z|) ({\bar{p}_x}^2-{\bar{p}_y }^2)\right. \right.\nonumber\\
&+& \left.\left.(q'^0p^0+|q'_z| |p_z|)(q^0 p'^0+|q_z| |p'_z|+{\bar{p}_x}^2-{\bar{p}_y }^2)\right] \right\} \sum_{s , s' =1}^{2} \sum_{r,r' =1}^{2} \hat{a}^\dagger_{s'}(p') \hat{a}_{s}(p) \hat{a}^\dagger_{r'}(q') \hat{a}_{r}(q) \nonumber\\
&\times&  \{[(\epsilon_r(q).\epsilon_s(p)) (\epsilon_{r'}(q').\epsilon_{s'}(p')) - (\epsilon_r(q).\epsilon_{s'}(p')) (\epsilon_{r'}(q').\epsilon_s(p))]\} ~ .\label{Hamiltoni}
\eea
Assuming $|\bar{p}_x|=|\bar{p}_y|$, we obtain
\bea
\hat{H}_I (t)&=&\frac{ g^2_{a\gamma \gamma} }{16}  \left(\frac{1}{2\pi \beta_\perp^2}\right)^2  \left(\frac{1}{2\pi \beta_{\|}^2}\right) \left(16\pi^2 \beta_\perp^6\right)^2\int \frac{dp_z}{(2\pi)^{3/2}} \int \frac{dp'_z}{(2\pi)^{3/2}} \int \frac{dq_z}{(2\pi)^{3/2}}\int \frac{dq'_z}{(2\pi)^{3/2}}\int \frac{d^3\bold{k}}{(2\pi)^{3}}  \frac{e^{i(p'^{0}-p^0+q^0-q'^0)t}}{\sqrt{p^0 q^0 p'^0 q'^0}} \nonumber\\
&\times& \int d^3\bold{x'} \int d^3\bold{x}~ ~e^{-2\beta_\perp^2[(x-W)^2+y^2]}~e^{-2\beta_\perp^2[x'^2+y'^2]}~\frac{i e^{-i\bold{k}.(\bold{x-x'})}}{(q^0-q'^0)^2-\bold{k}^2-m_a^2+ik^0\Gamma_\beta(k^0)}\nonumber\\
&\times&
~e^{\frac{-(q_z'-\bar{q}_z)^2}{4 \beta_\|^2}}  e^{\frac{-(p'_z-\bar{p}_z)^2}{4 \beta_\|^2}} e^{\frac{-(q_z-\bar{q}_z)^2}{4 \beta_\|^2}}  e^{\frac{-(p_z-\bar{p}_z)^2}{4 \beta_\|^2}}~e^{i(p'_z-p_z).(z-\bar{z})} ~e^{-i(q_z-q'_z).(z'-\bar{z}')}\nonumber\\
&\times&\left\{4 x' (x-W)(q'^0 p^0+|q'_z| |p_z|+q^0 p'^0+|q_z| |p'_z|+q^0 p^0+|q_z| |p_z|+p'^0 q'^0+|p'_z| |q'_z|-4 \bar{p}_x^2)+16\bar{p}_x^2 x' y\right. \nonumber\\ 
&+&\left.4 y'\left[y(q'^0 p^0+|q'_z| |p_z|+q^0 p'^0+|q_z| |p'_z|+q^0 p^0+|q_z| |p_z|+p'^0 q'^0+|p'_z| |q'_z|+4 \bar{p}_x^2)+4\bar{p}_x^2 (W-x)\right] \right. \nonumber\\
&+&\left. \frac{2 i}{\beta_\perp^2}\left[-(q^0 p^0+|q_z| |p_z|-p'^0 q'^0-|p'_z| |q'_z|) \left[\bar{p}_x(W-x-x'+y-y') \right]\right. \right. \nonumber\\
&+& \left. \left. (q'^0 p^0+|q'_z| |p_z|-q^0 p'^0-|q_z| |p'_z|)\left[\bar{p}_x (W-x+x'+y+y') \right]\right]  \right. \nonumber\\
 &+& \left.\frac{1}{\beta_\perp^4} \left[-(p^0q^0+|p_z| |q_z|) (p'^0 q'^0+|p'_z| |q'_z|)
+(q'^0p^0+|q'_z| |p_z|)(q^0 p'^0+|q_z| |p'_z|)\right] \right\}  \nonumber\\
&\times&\sum_{s , s' =1}^{2} \sum_{r,r' =1}^{2} \{[(\epsilon_r(q).\epsilon_s(p)) (\epsilon_{r'}(q').\epsilon_{s'}(p')) 
- (\epsilon_r(q).\epsilon_{s'}(p')) (\epsilon_{r'}(q').\epsilon_s(p))]\} ~ \hat{a}^\dagger_{s'}(p') \hat{a}_{s}(p) \hat{a}^\dagger_{r'}(q') \hat{a}_{r}(q)~,\label{Hamiltoni}
\eea
and after integrating over axion momentum, we find
\bea
\hat{H}_I (t)&\simeq&\frac{ g^2_{a\gamma \gamma} }{16} \left(\frac{1}{2\pi \beta_\perp^2}\right)^2 \left(\frac{1}{2\pi \beta_{\|}^2}\right)
 \left(16\pi^2 \beta_\perp^6\right)^2 \int \frac{dp_z}{(2\pi)^{3/2}} \int \frac{dp'_z}{(2\pi)^{3/2}}\int \frac{dq_z}{(2\pi)^{3/2}}\int \frac{dq'_z}{(2\pi)^{3/2}} \frac{e^{i(p'^{0}-p^0+q^0-q'^0)t}}{\sqrt{p^0 q^0 p'^0 q'^0}} \nonumber\\
&\times& \int_{-\infty}^\infty dz ~\int_{-\infty}^\infty dz' ~ \int d{\bold{x_\perp}^2} \int d{\bold{x'}^{2}_\perp}~e^{-2\beta_\perp^2[(x-W)^2+y^2]}~e^{-2\beta_\perp^2[x'^2+y'^2]}~\frac{e^{-m_a\sqrt{(\bold{x}_\perp - \bold{x_\perp'})^2+ (z-z')^2}}}{\sqrt{(\bold{x}_\perp - \bold{x_\perp'})^2+ (z-z')^2}}\nonumber\\
&\times&e^{\frac{-(q_z'-\bar{q}_z)^2}{4 \beta_\|^2}}  e^{\frac{-(p'_z-\bar{p}_z)^2}{4 \beta_\|^2}} e^{\frac{-(q_z-\bar{q}_z)^2}{4 \beta_\|^2}}  e^{\frac{-(p_z-\bar{p}_z)^2}{4 \beta_\|^2}}~e^{i(p'_z-p_z).(z-\bar{z})} ~e^{-i(q_z-q'_z).(z'-\bar{z}')}\nonumber\\
&\times&\left\{4 x' (x-W)(q'^0 p^0+|q'_z| |p_z|+q^0 p'^0+|q_z| |p'_z|+q^0 p^0+|q_z| |p_z|+p'^0 q'^0+|p'_z| |q'_z|-4 \bar{p}_x^2)+16\bar{p}_x^2 x' y\right. \nonumber\\ 
&+&\left.4 y'\left[y(q'^0 p^0+|q'_z| |p_z|+q^0 p'^0+|q_z| |p'_z|+q^0 p^0+|q_z| |p_z|+p'^0 q'^0+|p'_z| |q'_z|+4 \bar{p}_x^2)+4\bar{p}_x^2 (W-x)\right] \right. \nonumber\\
&+&\left. \frac{2 i}{\beta_\perp^2}\left[-(q^0 p^0+|q_z| |p_z|-p'^0 q'^0-|p'_z| |q'_z|) \left[\bar{p}_x(W-x-x'+y-y') \right]\right. \right. \nonumber\\
&+& \left. \left. (q'^0 p^0+|q'_z| |p_z|-q^0 p'^0-|q_z| |p'_z|)\left[\bar{p}_x (W-x+x'+y+y') \right]\right]  \right. \nonumber\\
 &+&  \left. \frac{1}{\beta_\perp^4} \left[-(p^0q^0+|p_z| |q_z|) (p'^0 q'^0+|p'_z| |q'_z|)
+(q'^0p^0+|q'_z| |p_z|)(q^0 p'^0+|q_z| |p'_z|)\right] \right\} \nonumber\\
&\times& \sum_{s , s' =1}^{2} \sum_{r,r' =1}^{2}\{[(\epsilon_r(q).\epsilon_s(p)) (\epsilon_{r'}(q').\epsilon_{s'}(p')) - (\epsilon_r(q).\epsilon_{s'}(p')) (\epsilon_{r'}(q').\epsilon_s(p))]\}~\hat{a}^\dagger_{s'}(p') \hat{a}_{s}(p) \hat{a}^\dagger_{r'}(q') \hat{a}_{r}(q)~. \label{axion-hamiltonian} 
\eea
In the context of a discrete approach, the discrete decomposition of the Gaussian distribution leads to the Hamiltonian for axions. After performing some calculations and substituting $\beta_\perp=1/\sigma$, we arrive at the following expression
\bea
\hat{H}_I (t)&\simeq&\frac{ g^2_{a\gamma \gamma} }{16} \int \frac{dp_z}{(2\pi)^{3/2}} \int \frac{dp'_z}{(2\pi)^{3/2}}~ \int \frac{dq_z}{(2\pi)^{3/2}}\int \frac{dq'_z}{(2\pi)^{3/2}}~\int_{-\infty}^\infty dz ~\int_{-\infty}^\infty dz' ~ \frac{e^{i(p'^{0}-p^0+q^0-q'^0)t}}{\sqrt{p^0 q^0 p'^0 q'^0}} \nonumber\\
&\times&\left(\frac{16\pi^4} {\sigma^4}\right) \left(\frac{1}{2\pi \beta_{\|}^2}\right)
~e^{\frac{-(q_z'-\bar{q}_z)^2}{4 \beta_\|^2}}  e^{\frac{-(p'_z-\bar{p}_z)^2}{4 \beta_\|^2}} e^{\frac{-(q_z-\bar{q}_z)^2}{4 \beta_\|^2}}  e^{\frac{-(p_z-\bar{p}_z)^2}{4 \beta_\|^2}}~e^{i(p'_z-p_z).(z-\bar{z})} ~e^{-i(q_z-q'_z).(z'-\bar{z}')}\nonumber\\
&\times& \sum_{s , s' =1}^{2} \sum_{r,r' =1}^{2}\{[(\epsilon_r(q).\epsilon_s(p)) (\epsilon_{r'}(q').\epsilon_{s'}(p')) - (\epsilon_r(q).\epsilon_{s'}(p')) (\epsilon_{r'}(q').\epsilon_s(p))]\}~ \hat{a}^\dagger_{s'}(p') \hat{a}_{s}(p) \hat{a}^\dagger_{r'}(q') \hat{a}_{r}(q)\nonumber\\
&\times&\left\{4{\sigma^2(q'^0 p^0+|q'_z| |p_z|+q^0 p'^0+|q_z| |p'_z|+q^0 p^0+|q_z| |p_z|+p'^0 q'^0+|p'_z| |q'_z|)}\left[\frac{ 0.336~ e^{-m_a \sqrt{W^2+ (z-z')^2}}}{\sqrt{W^2+ (z-z')^2}}\right.\right.\nonumber\\
&+&\left. \left.\frac{0.036~ e^{-m_a \sqrt{(W-\sigma)^2+ (z-z')^2}}}{\sqrt{(W-\sigma)^2+ (z-z')^2}} +\frac{0.036~ e^{-m_a \sqrt{(W+\sigma)^2+ (z-z')^2}}}{\sqrt{(W+\sigma)^2+ (z-z')^2}}-\frac{0.002~ e^{-m_a \sqrt{(W-2\sigma)^2+ (z-z')^2}}}{\sqrt{(W-2\sigma)^2+ (z-z')^2}}\right.\right.\nonumber\\
&-&\left.\left.\frac{0.002~ e^{-m_a \sqrt{(W+2\sigma)^2+ (z-z')^2}}}{\sqrt{(W+2\sigma)^2+ (z-z')^2}}+\frac{0.073~ e^{-m_a \sqrt{W^2+\sigma^2+ (z-z')^2}}}{\sqrt{W^2+\sigma^2+ (z-z')^2}}-\frac{0.004~ e^{-m_a \sqrt{W^2+4\sigma^2+ (z-z')^2}}}{\sqrt{W^2+4\sigma^2+ (z-z')^2}}\right.\right.\nonumber\\
&-&\left.\left.\frac{0.036~ e^{-m_a \sqrt{W^2+5\sigma^2-2W\sigma+ (z-z')^2}}}{\sqrt{W^2+5\sigma^2-2W\sigma+ (z-z')^2}}-\frac{0.036~ e^{-m_a \sqrt{W^2+5\sigma^2+2W\sigma+ (z-z')^2}}}{\sqrt{W^2+5\sigma^2+2W\sigma+ (z-z')^2}}\right.\right.\nonumber\\
&-&\left.\left.\frac{0.036~ e^{-m_a \sqrt{W^2+5\sigma^2-4W\sigma+ (z-z')^2}}}{\sqrt{W^2+5\sigma^2-4W\sigma+ (z-z')^2}}-\frac{0.036~ e^{-m_a \sqrt{W^2+5\sigma^2+4W\sigma+ (z-z')^2}}}{\sqrt{W^2+5\sigma^2+4W\sigma+ (z-z')^2}}\right.\right.\nonumber\\
&-&\left.\left.\frac{0.164~ e^{-m_a \sqrt{W^2+8\sigma^2-4W\sigma+ (z-z')^2}}}{\sqrt{W^2+8\sigma^2-4W\sigma+ (z-z')^2}}-\frac{0.164~ e^{-m_a \sqrt{W^2+8\sigma^2+4W\sigma+ (z-z')^2}}}{\sqrt{W^2+8\sigma^2+4W\sigma+ (z-z')^2}}\right]\right.\nonumber\\
&+&\left.{4 p_x^2\sigma^2} \left[\frac{0.146~ e^{-m_a \sqrt{(W-\sigma)^2+ (z-z')^2}}}{\sqrt{(W-\sigma)^2+ (z-z')^2}} +\frac{0.146~ e^{-m_a \sqrt{(W+\sigma)^2+ (z-z')^2}}}{\sqrt{(W+\sigma)^2+ (z-z')^2}}+\frac{0.664~ e^{-m_a \sqrt{(W-2\sigma)^2+ (z-z')^2}}}{\sqrt{(W-2\sigma)^2+ (z-z')^2}}\right.\right.\nonumber\\
&+&\left.\left.\frac{0.664~ e^{-m_a \sqrt{(W+2\sigma)^2+ (z-z')^2}}}{\sqrt{(W+2\sigma)^2+ (z-z')^2}}-\frac{0.292~ e^{-m_a \sqrt{W^2+\sigma^2+ (z-z')^2}}}{\sqrt{W^2+\sigma^2+ (z-z')^2}}-\frac{1.328~ e^{-m_a \sqrt{W^2+4\sigma^2+ (z-z')^2}}}{\sqrt{W^2+4\sigma^2+ (z-z')^2}}\right.\right.\nonumber\\
&-&\left.\left.\frac{0.146~ e^{-m_a \sqrt{W^2+5\sigma^2-2W\sigma+ (z-z')^2}}}{\sqrt{W^2+5\sigma^2-2W\sigma+ (z-z')^2}}-\frac{0.146~ e^{-m_a \sqrt{W^2+5\sigma^2+2W\sigma+ (z-z')^2}}}{\sqrt{W^2+5\sigma^2+2W\sigma+ (z-z')^2}}\right.\right.\nonumber\\
&+&\left.\left.\frac{0.146~ e^{-m_a \sqrt{W^2+5\sigma^2-4W\sigma+ (z-z')^2}}}{\sqrt{W^2+5\sigma^2-4W\sigma+ (z-z')^2}}+\frac{0.146~ e^{-m_a \sqrt{W^2+5\sigma^2+4W\sigma+ (z-z')^2}}}{\sqrt{W^2+5\sigma^2+4W\sigma+ (z-z')^2}}\right]\right\}~.
  \label{H-axion-final}
\eea
Since the points ${\bar{z}}$ and ${\bar{z}'}$ range from $0$ to $L$, the integrals over the momenta of the $z$ component yield the rectangular function \(\Pi_0^L(z)\).
\section{Determining Quantum Fisher Information} \label{appendix-QFI}
From the Hamiltonian of Eq. \eqref{AHamilton}, the $\chi_a$ factor of the unitary operator ($\hat{U} = \exp[i \chi_a~ \hat{a}^\dagger \hat{a}^\dagger \hat{a} \hat{a}]$) reads
\bea
\chi_a=\frac{2048 \pi^6 g'^2_{a\gamma \gamma} \mathcal{F} c^2}{L \sigma^2}~ \mathcal{B}(\sigma)~,
\eea
where $\mathcal{B}(\sigma)$ is expressed in Eq. \eqref{B}. According to \cite{Zain:2023}, we consider the SQV states as the initial states
\bea
\ket{\text{SQV}}=\hat{U}_{sq}(\xi) \ket{0}=e^{(\xi^\ast \hat{a} \hat{a} - \xi \hat{a}^\dagger \hat{a}^\dagger)/2} \ket{0}~, \label{sqv state}
\eea 
where $\xi=re^{i\theta}$ is the squeezing parameter. Therefore
\bea
 \ket{\psi_0}=\ket{\text{SQV}}~.
\eea
The one-mode SQV state in the number basis is defined as \cite{Lvovsky:2016}
\bea
\ket{\text{SQV}}=\frac{1}{\sqrt{\cosh(r)}} \sum_{n=0}^\infty \left(-e^{i\phi} \tanh(r)\right)^n \frac{\sqrt{(2n)!}}{2^n n!} \ket{2n}~,
\eea
\bea
\ket{\text{SQV}}= \sum_{n=0}^\infty C_n \ket{2n}~ ,
\eea
where 
\bea
C_n=\frac{1}{\sqrt{\cosh(r)}}  \left(-e^{i\phi} \tanh(r)\right)^n \frac{\sqrt{(2n)!}}{2^n n!} ~.
\eea
It can be demonstrated that, for the SQV state 
\bea
\braket{N}_{\text{SQV}}=\braket{\hat{a}^\dagger \hat{a}}=\text{sinh}^2(r)~.
\eea
Therefore the final state can be achieved by applying the unitary operator on the SQV state
\bea
\ket{\psi}=\hat{U} \ket{\psi_0}~.
\eea
The QFI associated with any parameter $u$ can be expressed as
\bea
F_{u}=4\, \text{Re} \left(\braket{\partial_u \psi |\partial_u \psi} - \braket{\partial_u \psi | \psi}\braket{\psi | \partial_u \psi}\right)~.
\eea
After performing calculation of ${\chi_a}$ parameter using the initial SQV state, we obtain
\bea
{F}_{\chi_a}= 4\left(2 \cosh^6(r) ~\sinh^2(r) + 44 \cosh^4(r)~ \sinh^4(r) + 50 \cosh^2(r) ~\sinh^6(r)\right)~, \label{fisher matrix}
\eea
where using $\braket{N} = \sinh^2(r)$, we have
\bea
{F}_{\chi_a} = 8N(48 N^3 + 72 N^2 + 25N + 1)~.\label{Fxa}
\eea
\section{Optimal quantum fisher information}
We repeat deriving the QFI matrix for both the initial coherent state and the squeezed coherent (SQC) state. By comparing the results with those obtained from the initial SQV state, one can determine the optimal Fisher matrix. For the coherent state, we have
\bea
\ket{\alpha}=\mathcal{D}(\alpha) \ket{0}~,
\eea
where
\bea
\mathcal{D}(\alpha)= e^{(-\alpha \hat{a}^\dagger -\alpha^\ast \hat{a})}~,
\eea
is the displacement operator. The coherent state in the number basis is
\bea
\ket{\alpha}=C_n \ket{n}=e^{-\frac{|\alpha|^2}{2}} \sum_{n=0}^{\infty} \frac{\alpha^n}{n!} \ket{n}=e^{-\frac{|\alpha|^2}{2}}e^{(-\alpha \hat{a}^\dagger -\alpha^\ast \hat{a})} \ket{0}~.
\eea
The QFI matrix for the coherent state gives
\bea
{F}_{\chi_a}=8 N(2 N^2+N)~, \label{sb} 
\eea
where
 $\braket{N}_{C}=\braket{\hat{a}^\dagger \hat{a}}=|\alpha|^2$.
For the SQC state, we have
\bea
\ket{\alpha , \xi}=\mathcal{D}(\alpha) \ket{\text{SQV}}~,
\eea
where $\xi$ is the squeezing parameter and $\ket{\text{SQV}}$ is defined as in \eqref{sqv state}. The state $\ket{\alpha , \xi}$ in the number basis is
\bea
\ket{\alpha,\xi}=C_n \ket{n}=e^{\left(-\frac{1}{2}|\alpha|^2 -\frac{1}{2} \alpha^{\ast2}  e^{i\theta} \tanh(r)\right)}[n!~ \cosh(r)]^{-\frac{1}{2}}  [\frac{1}{2} e^{i \theta} \tanh(r)]^{\frac{n}{2}}H_n \left(\gamma [e^{i \theta} \sinh(2r)]^{-\frac{1}{2}}\right)\ket{n} ~,
\eea
where $H_n(x)$ are Hermite polynomials and
\bea
\gamma= \alpha \cosh(r)+ \alpha^\ast e^{i\theta} \sinh(r)~.
\eea
We can show 
 $\braket{N}_{\text{SQC}}=\braket{\hat{a}^\dagger \hat{a}}=|\alpha|^2+\text{sinh}^2  r$.
After some calculations, we find that for the SQC state the QFI becomes
\bea
{F}_{\chi_a}&=&4 \left( 2 \text{cosh}^6(r)~ \text{sinh}^2(r)+44 \text{cosh}^4(r) ~ \text{sinh}^4(r) +50 \text{cosh}^2(r) ~ \text{sinh}^6(r)   +16 |\alpha|^2 \text{sinh}^6(r)+44 |\alpha|^2 \text{ sinh}^2(r) ~ \text{cosh}^4(r)\right. \nonumber\\
&&\left. + 156 |\alpha|^2 \text{ sinh}^4(r) ~ \text{cosh}^2(r)+63 |\alpha|^4 \text{ sinh}^2(r)~ \text{cosh}^2(r)  +18 |\alpha|^4 \text{ sinh}^4(r)+4 |\alpha|^6 (\text{ sinh}^2(r)+ \text{cosh}^2(r)) \right.  \nonumber\\
&&\left. + 2 \text{ sinh}^2(r) ~ \text{cosh}^2(r)(\alpha^4 e^{-2i \theta}+ \alpha^{\ast4} e^{2i\theta}) - (\alpha^2 e^{-i \theta}+ \alpha^{\ast2} e^{i\theta})\left[ 48  \text{ sinh}^3(r) ~ \text{cosh}^3(r) + 34  \text{ sinh}^5(r) ~ \text{cosh}(r)\right.\right. \nonumber\\
&&\left. \left. + 2 \text{ sinh}(r) ~ \text{cosh}^5(r)  + 28 |\alpha|^2 \text{ sinh}^3(r)~ \text{cosh}(r) + 12 |\alpha|^2 \text{ sinh}(r) ~ \text{cosh}^3(r) + 4 |\alpha|^4 \text{ sinh}(r) ~ \text{cosh}(r)\right] \right)~.
\eea
Since for SQC state, $\braket{N}^4=|\alpha|^8+4|\alpha|^6\text{sinh}^2(r)+6|\alpha|^4\text{sinh}^4(r)+4|\alpha|^2\text{sinh}^6(r)+\text{sinh}^8(r)$, therefore
\bea
{F}_{\chi_a}&<&384{\braket{N}^4}~.
\eea
Finally, when comparing this QFI with that of the SQV state, as shown in Eq. \eqref{Fxa}, we find that it is smaller in the terms consist of $N^4$. In the limit where the photon number of the state is large, $N \gg 1$, we focus solely on the highest power of $N$ (i.e. $N^4$). Consequently, assuming that  $\braket{N}_{\text{SQC}}\simeq \braket{N}_{\text{SQV}}$, the SQC state does not enhance the QFI compared to the SQV state. Therefore, the QFI of the SQV state remains optimal.
\end{appendices}


\begingroup 
  \makeatletter
  \let\ps@plain\ps@empty
  \makeatother
  \bibliography{bibl}
\endgroup

\end{document}